\documentclass[twocolumn,aps,superscriptaddress]{revtex4-1}

\usepackage{graphicx,amssymb,soul,color,amsmath,bm}
\usepackage{epstopdf,hyperref}
\usepackage{braket,dsfont}
\usepackage[mathcal]{euscript}

\hypersetup{colorlinks=true,citecolor=blue,linkcolor=magenta}

\sethlcolor{yellow}
\allowdisplaybreaks

\begin{document}

\title{Beyond perturbation theory: A time-dependent approach to inelastic scattering spectroscopies in- and away from equilibrium}
\author{Krissia Zawadzki}
\affiliation{Department of Physics, Northeastern University, Boston, Massachusetts 02115, USA}
\author{Luhang Yang}
\affiliation{Department of Physics, Northeastern University, Boston, Massachusetts 02115, USA}
\author{Adrian E. Feiguin}
\affiliation{Department of Physics, Northeastern University, Boston, Massachusetts 02115, USA}

\date{\today}
\begin{abstract}
We propose a non-perturbative numerical approach to calculate the spectrum of a many-body Hamiltonian with time and momentum resolution by exactly recreating a scattering event using the time-dependent Schr\"odinger equation. Akin an actual inelastic scattering experiment, we explicitly account for the incident and scattered particles ({\it e.g.} photons, neutrons, electrons...) in the Hamiltonian and obtain the spectrum by measuring the energy and momentum lost by the particle after interacting with the sample. 
We illustrate the method by calculating the spin excitations of a Mott-insulating Hubbard chain after a sudden quench with the aid of the time-dependent density matrix renormalization group (tDMRG) method. 
Our formalism can be applied to different forms of spectroscopies, such as neutron and Compton scattering, and electron energy-loss spectroscopy (EELS), for instance.

\end{abstract}

\maketitle

\section{Introduction}
Inelastic scattering or, in general, energy-loss spectroscopies, are an exceptional tool that enable experimentalists to peek into the hidden mechanisms responsible for the magnetic and electronic excitations inside solids and molecules. For instance, the inelastic neutron scattering cross section is proportional to the magnetic dynamical structure factor,  while Compton and energy-loss (EELS) spectra are related to the charge density excitations. \cite{eels1,eels2,eels3,eels4}. In all these cases, a sample is subject to a beam of incident particles (neutrons, X-ray photons, and electrons, respectively, in the aforementioned cases). As their names imply, these techniques rely on analyzing the energy distribution of the scattered particles after they have interacted with the sample. In most cases, particles are able to penetrate several atomic layers before they are reflected, transferring part of their energy and momentum to the degrees of freedom in the specimen in the process. The corresponding information is gathered by measuring the energy and momentum ``lost'', that correspond, by conservation, to the energy and momentum transferred to the solid. 

The foundations of time-dependent perturbation theory for quantum scattering are due to Schwinger and Lippman  \cite{Lippmann1950} , who derived an expression for the scattering cross-section as a linear response that accounts from the transition rate between the eigenstates as in Fermi's Golden Rule.  If the system originally is in the ground state $|0\rangle$, this approach allows one to express the energy and momentum resolved spectral function as (we use units in which $\hbar=1$): 
\begin{equation}
S_O(\mathbf{k},\omega) = 2\pi \sum_{n} |\langle n|O_\mathbf{k}|0\rangle|^2 \delta(\omega-E_n+E_0),
\label{gamma}
\end{equation}
where $\mathbf{k}$ represents the momentum quantum number, $\ket{n}$ are the eigenstates of the system's unperturbed with energy $E_n$, and $O_\mathbf{k}$ is the Fourier transform of the operator $O$ associated to the interaction potential between the incident particles and the degrees of freedom inside the sample (spin or electron density), which typically enters as a local contact term, as we describe in the next section. 

The relative simplicity of the previous expression has allowed theorist and experimentalists to model and compare predictions with theory very accurately. In the particular context of strongly correlated quantum matter, these calculations are carried out by means of state of the art computational techniques. These include: exact diagonalization\cite{Dagotto1994}, which is limited to small system sizes;
quantum Monte Carlo, that is conditioned by the sign problem and requires uncontrolled analytic continuations and the use of the max entropy approximation\cite{Schuttler1986,Sandvik1998,Silver1990,Gubernatis1991,Syljuaasen2008,Fuchs2010,Sandvik2016,Shao2017}; dynamical density matrix renormalization group (DMRG) \cite{Hallberg1995,Kuhner1999,Jeckelmann2002}, which is  computationally demanding and applies mostly to quasi one-dimensional systems; the time-dependent DMRG \cite{Daley2004,White2004a,Feiguin2005,vietri,Feiguin2013b,PAECKEL2019} and recent variations using Chebyshev expansions \cite{Holzner2011,Wolf2015,Xie2018}, also limited by the entanglement growth. In addition, matrix product states have been used to build variational forms for excited states\cite{Vanderstraeten2015,Vanderstraeten2015b}. Similar ideas were explored with variational Monte Carlo, that can be easily extended to higher dimensions and are free from the sign problem\cite{Li2010,DallaPiazza2014,Ferrari2018,Hendry2020}.

Despite their success, these methods hit a hard wall when it comes to studying dynamics of a system far from equilibrium, as a result of a pump or a quench, for instance. In that case, it is appropriate to assume that the system is initially in 
a generic state $|\phi\rangle=\sum_n a_n |n\rangle$. The expression for the spectral function is now given as\cite{Zawadzki2019}: 
\begin{equation}
S_O(k,\omega,t) = 4\pi^2\sum_m \left| \sum_n a_n \langle m|O|n\rangle \delta^{t}(\omega-\omega_{mn}) \right|^2,
\label{Skw}
\end{equation}
where we have introduced the time dependence in the definition 
\begin{equation}
    \delta^t(\omega)=\frac{1}{\pi}\frac{\sin{(\omega t/2)}}{\omega}\underset{t\rightarrow \infty}{\rightarrow} \delta(\omega)
\end{equation}
Unlike the equilibrium case, this expression cannot be simplified and, at the same time, most methods listed above no longer apply \cite{Freericks2009,Shao2016,Zawadzki2019}, forcing us to rely on the limited power of exact diagonalization.

The aim of this work is computing the spectrum of energy-loss spectroscopies without resorting to perturbation theory nor to the calculation of the full eigen-spectrum of the system. Working directly in the time-domain, we propose to simulate the entire scattering event by solving the time-dependent Schr\"odinger equation for an equivalent system comprising the sample, a source and a detector. The interaction terms between incident and reflected particles are included explicitly and a response function for the detector can be calculated exactly in real time. In this scenario, the spectrum can be conveniently obtained with low computational effort using the time-dependent DMRG (tDMRG). Besides the obvious numerical advantages, our method is able to reveal features in the scattering spectrum that remain hidden in the conventional perturbative expression obtained from linear response. 

Our paper is organized as follows: In section \ref{sec:methods}, we present the mathematical formulation and the numerical scheme used to simulate the scattering event using the time-dependent DMRG method. In section \ref{sec:results}, we show numerical results for the Heisenberg chain and the Hubbard chain, in- and away from equilibrium. We finally close with a discussion.   

\section{Method} \label{sec:methods}

\begin{figure}
        \centering
        \includegraphics[width=\columnwidth]{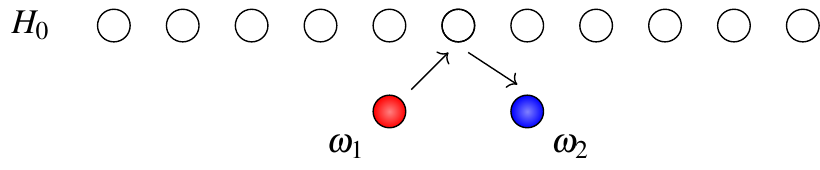}
        \caption{Scattering process of a single particle in 1D.}
        \label{fig:energy_loss_spectroscopy}
    \end{figure}

    In ``energy loss'' spectroscopies an incident particle (photon, neutron, electron) with initial energy $\omega_s$ interacts with a system described by a Hamiltonian $H_0$ and is inelastically reflected with final energy 
    $\omega_d$, typically off resonance. In an actual experiment, the energies $\omega_s=k_s^2/2m$ and $\omega_d=k_2^2/2m$ correspond to the kinetic energy of free particles (neutrons, electrons) in the beam (obviously, these expression do not apply to photons). Conservation laws imply that the energy lost by the particle has been transferred to the system $\Delta E = \omega_d - \omega_s$; see Fig. \ref{fig:energy_loss_spectroscopy}.
    As mentioned in the introduction, the measurement of the cross section of the outgoing particle is directly related to the excitation spectrum of the sample.

    To model this process we consider the Hamiltonian 
    \begin{align}
        \label{eq:Hfull}
        H = H_0 + H_d + V,
    \end{align}
    where $H_0$ is the Hamiltonian for the system of interest characterized by the energy scale $J$ and
    \begin{align}
        \label{eq:H_d}
        H_d = \omega_s ns + \omega_d n_d
    \end{align}
    represents the energy of and incoming particle with energy $\omega_s$ and outgoing with energy $\omega_d$. From now on we will refer to the ``orbitals'' representing these two states as ``source'' and ``detector''/``probe'', respectively. The term $V$ is a ``contact'' interaction between the particles and the sample that remains to be determined, depending on the nature of the spectroscopy of interest. 

For simplicity, let us first focus on the energy spectrum without momentum resolution. In this case, the contact term acts only on a site that we label as ``0''.   
    We assume that there is no absorption and the incident particle can only be reflected. 
    We want to represent a single particle scattering event, in which initially orbital $s$ is occupied, while $d$ is empty. The term $V$ is responsible for making the particle undergo a transition from a state with energy $\omega_s$ to a state with final energy $\omega_d$, due to either the Coulomb interaction or some other effect. In the case the force is of electrostatic origin, the potential is described as:
    \begin{eqnarray}
        V & = & J' n_0 (n_s+n_d)  (c_s^\dagger c_d + H.c.) \nonumber \\
        & = & J' n_0 (c_s^\dagger c_d + H.c.),
        \label{eq:V}
    \end{eqnarray}
    where $n_0$, $n_s$ and $n_d$ are the occupation numbers of the system's orbital ``0'', source and probe, respectively.
    Since the particle can only be in the source or in the detector, $n_s + n_d = 1$ is a constraint and must be satisfied at all times. Note that the creation and annihilation operators in this expression can be either bosonic or fermionic since there is only one such particle and its nature does no play a role. The constant $J'$ is a matrix element that will depend on the particular details of the electronic wave function and is assumed to be small.
    
    \begin{figure}
        \centering
         \includegraphics[width=\columnwidth]{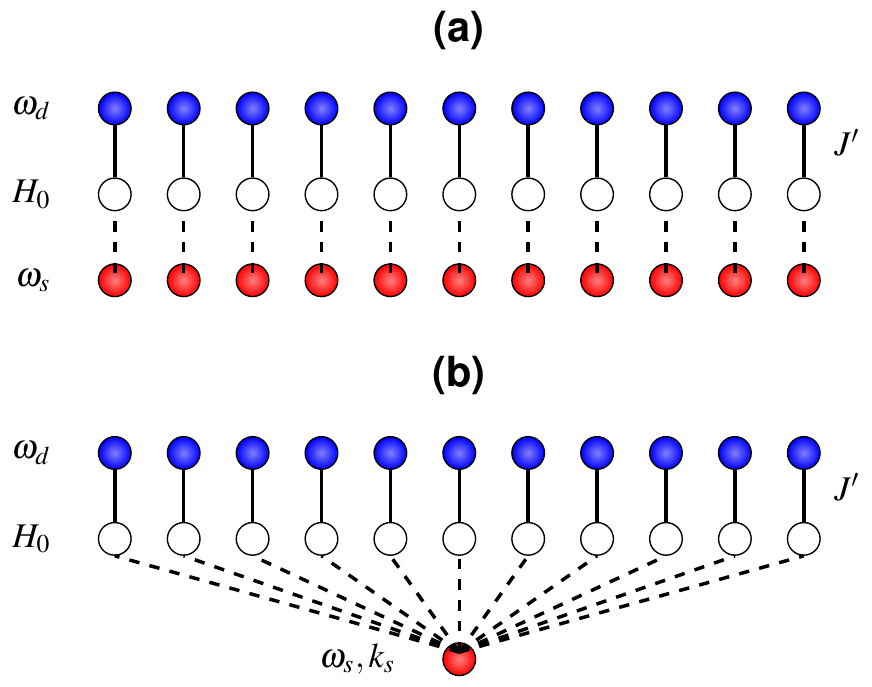}
        \caption{Possible geometries used in simulating a scattering event of a single particle with a 1D sample: (a) chain geometry in real space  and (b) star in momentum space. The empty circles represent the sample orbitals, while red (blue) filled circles indicate the source(detector) orbitals for the incident and scattered particle, respectively. Orbitals connected by lines interact via the perturbation $V$.}
        \label{fig:energy_loss_spectroscopy_k}
    \end{figure}

    For the case of neutrons interacting via (longitudinal) spin interactions, the perturbation can be expressed as
    \begin{align}
        \label{eq:Vspin}
        V = & J' S^z_0  (c_s^\dagger c_d + H.c.),
    \end{align}
    where,  the constraint is now $S^z_s + S^z_d = 1/2$. Note that the scenario in which the probe particles are photons requires more care, because it involves the creation and annihilation of particles\cite{Zawadzki2020}.
    
    We now have all the ingredients to measure the energy loss by the particle after the scattering event. 
    At time $t=0$, we consider the total wave function of the system is 
    \begin{align}
        \label{eq:psi_0}
        \ket{\Psi(t=0)} = \ket{\phi}\otimes \ket{n_s=1}\otimes \ket{n_d=0},
    \end{align}
    where $\ket{\phi}$ is the state of the sample (in or away from equilibrium), and $|n_s\rangle$ and $|n_d\rangle$ describe the states for source and detector, respectively.

    Then, the coupling $J'$ is turned on and full the system $H$ in time. The occupation of the detector $\langle n_d(t) \rangle$ will be proportional to the spectral density at energy $\omega_d$, as $S_O(\omega_d) \propto \lim_{t\rightarrow \infty} n_d(t)/t$. One can easily show (see Appendix \ref{apx:PT}) that, in the limit $J'<< J$,  one recovers the same result obtained from perturbation theory. As we shall discuss later in the implementation, three important details require special attention: (i) the full spectrum is only recovered after scanning $\omega_d$ over an energy range; (ii) in our scheme with just one source and probe orbitals, at sufficiently long times, the particle oscillates back and forth between the two. Hence, the $t\rightarrow \infty$ limit is not well defined. Finally, (iii) since the treatment of the interaction does not rely on perturbation theory, the measure $\langle n_d(t) \rangle$ will contain all contributions to all orders. 
    
    \subsection{Momentum resolution}
    To adapt the previous ideas to translational invariant systems, we now model the source and detector to account for the momentum of the incoming and outgoing particle.  For illustration and simplicity, we limit our discussion to the one-dimensional case, but the same considerations can be generalized to any geometry. Below, we present two alternative but equivalent forms that will yield similar results, but will differ in their implementation.
    
    \begin{figure}
    \centering
    \includegraphics[width=0.95\columnwidth]{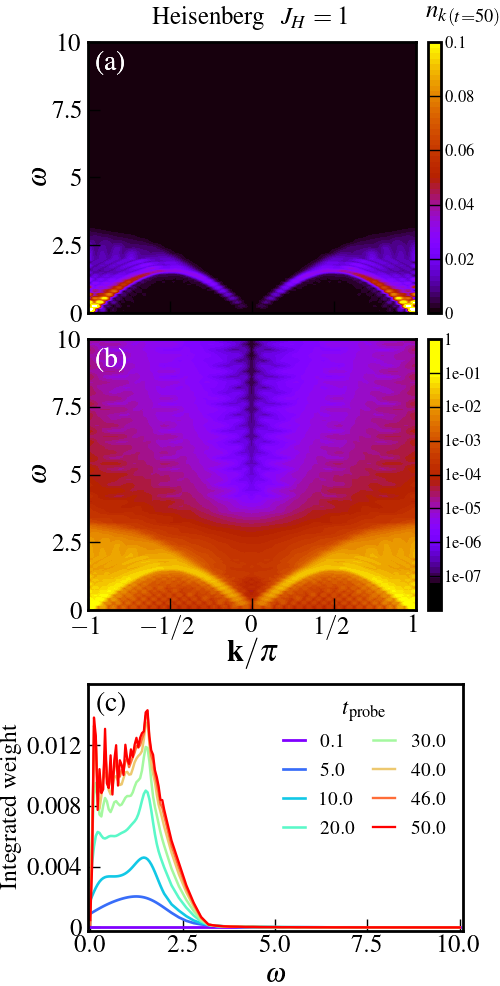}

    \caption{Momentum-resolved neutron scattering spectrum of Heisenberg chains of size $L=32$ at the final time $t_{probe} J = 50$. Color scales are (a) linear and (b) logarithmic. Integrated weight is depicted in (c)}
    \label{fig:spec_Heisenberg}
\end{figure}

    \subsubsection{Chain geometry}

    We first describe what we refer to as the ``chain geometry''\cite{Zawadzki2019}: both the source and detector are represented as two parallel chains of orbitals with the same number of ``sites'' as the  the system under study, represented by Hamiltonian $H_0$, as shown in in Fig. \ref{fig:energy_loss_spectroscopy_k}(a). 
    
    The Hamiltonian $H_d$ is now given by
    \begin{align}
        \label{eq:Hdk}
        H_d & = \omega_s \sum_\ell n_{1\ell} + \omega_d \sum_\ell n_{2\ell}
    \end{align}
and the interaction between system and the source is written as
    \begin{align}
        \label{eq:Vk}
        V = J' \sum_\ell O_\ell (c_{1\ell}^\dagger c_{2\ell} + H.c.),
    \end{align}
where $O_\ell$ is some generic diagonal local operator acting on site $\ell$ (we consider $O_\ell=O^\dagger_\ell$, but the formalism can be generalized to other cases).    

    \begin{figure}
    \centering
        \includegraphics[width=0.95\columnwidth]{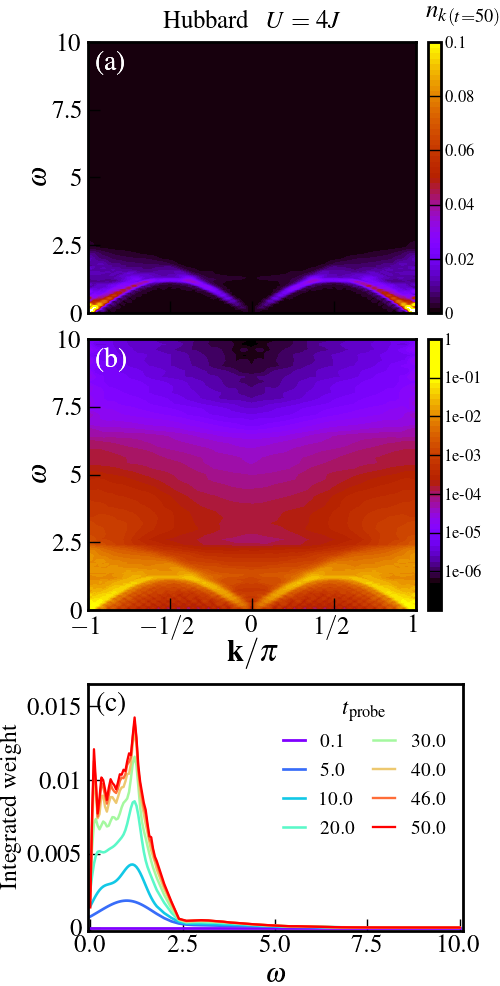}
        \caption{Neutron scattering spectrum of half-filled Hubbard chains of size $L=32$ with Coulomb coupling $U/J=4.0$. Momentum resolved spectrum at final time $tJ=50$ is plotted in (a) linear and (b) logarithmic color scale. Integrated weight  is depicted in (c).}

    \label{fig:spec_Hubbard_U=4.0_half}
\end{figure}

\begin{figure}
    \centering
    \includegraphics[width=0.95\columnwidth]{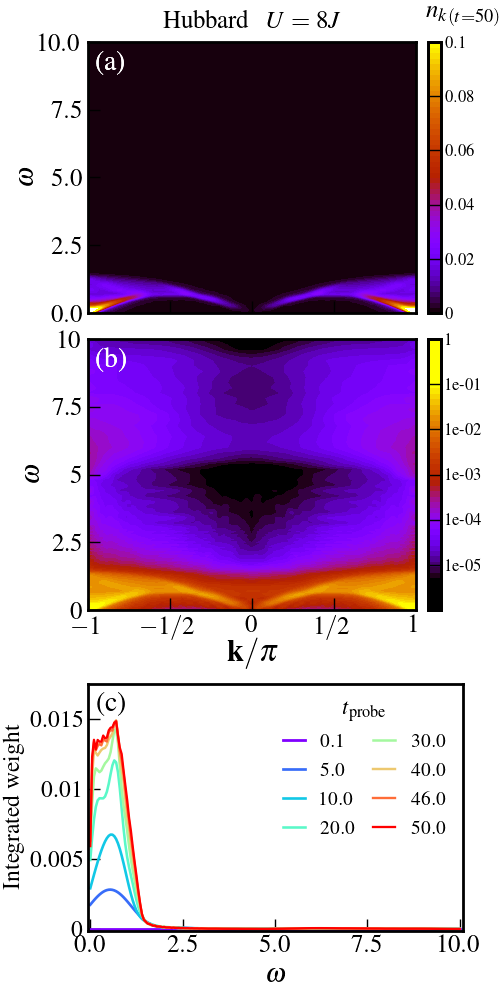}
    \caption{Same as figure \ref{fig:spec_Hubbard_U=4.0_half} but for $U/J = 8$.}
    \label{fig:spec_Hubbard_U=8.0_half}
\end{figure}

    Since source, probe and system are extended, and the interaction term is translational invariant, momentum conservation is ensured. 
    In the present setup, the initial state of the full system is given by
    \begin{align}
        \label{eq:psi0k}
        \ket{\Psi(t=0)} = & 
        \ket{\phi}\otimes \ket{n_s=1,k_s=k_0}\otimes \ket{n_d=0},
    \end{align}
    where $k_0$ is the momentum of the incident particle at the source.

    By measuring the momentum distribution at the detector $n_{dk}(t)$:
        \begin{align}
        \label{eq:n2k}
        n_{dk}(t) = \frac{1}{L} \sum_{\ell, \ell'} e^{ik(\ell-\ell)} \langle c_{d\ell}^\dagger c_{d\ell'} (t)
        \rangle,
    \end{align}
    we obtain the full spectrum of the system with both time and momentum resolution.

    \subsubsection{Star geometry}
    The number of degrees of freedom can be reduced considerably by accounting explicitly for the fact that the incident particle can only assume one allowed value of momentum $k_0$. In this case, instead of representing the source by a chain, we do it as a single orbital with energy $\omega_s$ and momentum $k_0$. Therefore the Hamiltonian $H_d$ becomes:
    \begin{equation}
          H_d  = \omega_s n_{sk_0} + \omega_d \sum_\ell n_{d\ell},
    \end{equation}
    and the interaction:
        \begin{eqnarray}
        V & = & \frac{J'}{\sqrt{L}} \sum_{\ell} O_\ell  \left(e^{ik_0\ell} c_{d,\ell}^\dagger c_{s,k_0} + H.c. \right),
    \end{eqnarray}
    with $O_k=1/L \sum_\ell e^{ik\ell}O_\ell$. The corresponding geometry is illustrated in Fig.\ref{fig:energy_loss_spectroscopy_k}(b). Notice that while the complexity of the problem has been greatly reduced, the Hamiltonian now contains long-range terms.
    
    Finally, we point out that, besides the two described approaches, there is yet a third possibility: a ``double star'' geometry in which the probe is ``tuned'' to detect only a scattered particle with fixed momentum $k_d$. In this case, we find that the interaction would be written as: 
        \begin{eqnarray}
        V & = & \frac{J'}{L} \sum_{\ell} O_\ell \left(e^{i(k_0-k_d)\ell} c_{d,k_d}^\dagger c_{s,k_0} + H.c. \right).
    \end{eqnarray}
    When using this scheme, one needs to carry out one calculation for each value of $k_d$, increasing the computational overhead by a factor of $L$.

%

\subsection{DMRG implementation}

    In order to recast these ideas into a practical numerical solver, we will describe an implementation in the context of a the time-dependent DMRG method. For this purpose, we consider a chain with $L$ sites coupled to two auxiliary chains $s$ and $d$ accounting for the source and detector/probe of neutrons/electrons/photons. The main advantage of this setup, compared to the star geometries, is that the Hamiltonian remains local and allows for a straightforward Suzuki-Trotter decomposition of the evolution operator (For details about tDMRG we direct the reader to \cite{White2004a,vietri,PAECKEL2019}). As examples, we shall present two cases for prototypical Hamiltonians $H_0$: The Hubbard chain is defined as: 
        \begin{eqnarray}
H_{Hubbard}&=&-J \sum_{i=1,\sigma}^{L-1} \left(c^\dagger_{i\sigma} c_{i+1\sigma}+\mathrm{h.c.}\right) + \nonumber \\
&+&U \sum_{i=1}^L 
\left(n_{i\uparrow}-\frac{1}{2}\right)\left(n_{i\downarrow}-\frac{1}{2}\right),
\label{Hubbard}
\end{eqnarray}
where $U$ and $J$ parametrize the on-site Coulomb interaction and the hopping, respectively (the symbol $t$ is reserved for the time variable). In the large $U/J$ limit, the charge fluctuations are suppressed, and only the spin degree of freedom remains. In this regime, the low energy physics is well described by the one-dimensional Heisenberg model:
    \begin{equation}
        H_{Heis}=J_H\sum_i \vec{S}_i \cdot \vec{S}_{i+1},
        \label{Heisenberg}
    \end{equation}
    where the operators $\vec{S}$ represent $S=1/2$ spins and $J_H \sim 4J^2/U$.
    
    Without loss of generality, we use a spinless fermion to represent the incident and scattered particle.

For ground-state calculations, we use the conventional DMRG method to initialize the system. To ensure that the source particle is in a state with well defined momentum and that the detector is empty, we include a projector $H_{k_0}=|k_0\rangle \langle k_0|$ and a large positive potential term in the detector.  Alternatively, the chain can be in a far from equilibrium state, resulting from a quench or a pump, for instance. In either case, before the measurement starts, the scattering term is always ``turned off'' with $J'=0$. At $t=0$ the source and probe are connected, and one can start measuring the momentum distribution on the detector chain. This procedure is carried out by keeping enough DMRG states to ensure a truncation error of the order of $10^{-6}$, corresponding to a block dimension up to $m=400$ in the worse cases. We typically run the simulations to times of the order of $t_{\text{probe}}=50$ (in unities of $J^{-1}$) for each energy $\omega_d$. This represents hundreds of simulations, but they are all carried out in parallel independently. 

\section{Results}\label{sec:results}

\subsection{Heisenberg chain}

As a control case study, we first calculate the spectrum of a spin chain Eq.(\ref{Heisenberg}) with $J_H=1$ as our unit of energy. The one-dimensional Heisenberg model does not realize long range order and the antiferromagnetic correlations decay algebraically. In addition, its excitations are spinons (domain walls) that carry spin 1/2. The spectrum is gapless and bounded from below by the des Cloizeaux-Pearson dispersion $\pi J/2|\sin{k}|$ \cite{Cloizeaux1962} and the upper boundary of the continuum is $\pi J|\sin{(k/2)}|$ \cite{Muller1981}. This physics is realized in a number of quasi one-dimensional magnets and the spinon excitations have been experimentally confirmed \cite{Zheludev2001,Stone2003,Kenzelmann2004,Kohno2009,Lake2010,Bouillot2011,Schmidiger2013,Schmidiger2013b,casola2013field,Mourigal2013,Blosser2017,Ward2017,Gannon2019,Yang2019,keselman2020}. Since spinons are not conventional Landau quasiparticles, the spectrum exhibits singularities at the edges instead of a coherent band or dispersion. We show results at time $t=50$ in Fig. \ref{fig:spec_Heisenberg}; panels (a) and (b) display the momentum resolved spin dynamical spectral function obtained with our approach in linear and log scales, respectively, while panel (c) shows the integrated weight. The oscillations in (c) are due to the high-resolution of the measurement that reveals finite-size effects, since we are considering a finite chain of length $L=32$ (In finite systems the spectrum is a collection of delta peaks).

\begin{figure}
    \centering
    \includegraphics[width=0.95\columnwidth]{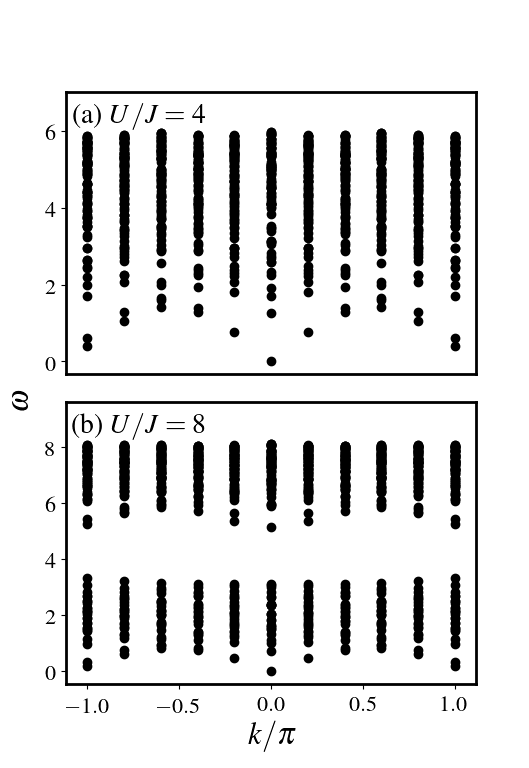} 
    \caption{Exact spectrum for a half-filled Hubbard chain with $L=10$ sites in the $Sz=0$ subspace for (a) $U/J=4$ and $(b) U/J=8$}
    \label{fig:ED}
\end{figure}

\begin{figure}
    \centering
    \includegraphics[width=0.95\columnwidth]{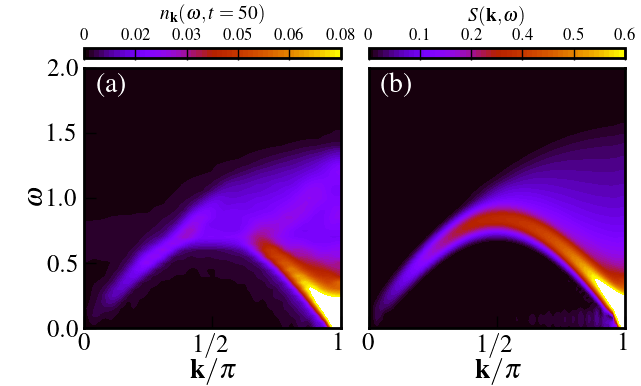}
    \caption{Comparison between (a) the non-perturbative time-dependent scattering approach introduced in this work and (b) equilibrium Green's function results for a Hubbard chain with $U/J=8$ using tDMRG (in arbitrary units). }
    \label{fig:spec_comparison}
\end{figure}

\subsection{Hubbard chain}

The Hubbard chain at half-filling is a Mott insulator with a charge gap that increases with $U/J$. However, spin excitations remain gapless and are also spinons, with a dispersion that resembles the one for the spin chain, but with a renormalized coupling $J_H \sim 4J^2/U$ \cite{EsslerBook}. Results of our calculations are shown in Figs.\ref{fig:spec_Hubbard_U=8.0_half} and \ref{fig:spec_Hubbard_U=4.0_half} for $U/J=8$ and 4, respectively. In both cases, the hopping $J=1$ is our unit of energy. We observe a well defined spinon spectrum  with a bandwidth determined by the renormalized value of $J_H$. However, in the $U/J=4$ case, an unexpected ``bubble`` of spectral weight is discerned above the continuum at energies near $\omega \sim 2.5$. These features are enhanced and clearly visible in the log scale plot, panel (b). Furthermore, the extra spectral weight can be appreciated in the integrated spectral density, panel (c). By paying further attention, we discover similar features in the $U/J=8$ results that, albeit being fainter than in the previous case, become obvious also in log scale, and occur at higher energies. This high energy bubble does not appear in calculations using linear response, Eq.(\ref{Skw}), begging us to try to understand its origin.

In order to identify the high-energy features, we resort to exact diagonalization calculations for small systems. In Fig.\ref{fig:ED} we show the eigenvalues for a chain with $L=10$ sites with total spin $S=0$ and (a) $U/J=4$ and (b) $U/J$ = 8. We observe that, besides the low energy manifold describing the spin physics traditionally associated to the Heisenberg limit, we also find a high energy manifold separated by a gap (the Mott gap). These states correspond to spin excitations in the upper Hubbard band. Why do they appear in our spectrum? To answer this question we recall that our formulation does not rely on perturbation theory and, therefore, it contains all contributions to $n_{dk}$ to all orders. 
Therefore, the appearance of the new features can be associated to high order contributions that, we should emphasize, are real in the sense that an idealized experimental setup with high resolution and no noise should be able to resolve them, particularly if the matrix elements (our $J'$) are large. However, despite this fact, this spectral weight is not associated to the spectral function (a quantity that arises from linear response), but to higher order transitions.   

It turns out that similar contributions can be observed in the low energy spectrum, as shown in Fig.\ref{fig:spec_comparison}. We here compare the results obtained by means of our scattering approach and the spectral function $S(k,\omega)$ obtained from equilibrium Green's functions using tDMRG\cite{White2004a,vietri,PAECKEL2019}. While the spectral function $S(k,\omega)$ displays a sharp lower edge, the higher order contributions are evident in the inelastic scattering spectrum with the appearance of a new ``branch'' in the middle of the continuum and a drop of spectral weight in the low energy edge of the spectrum between $k=0$ and $k\approx 2/3\pi$.

\begin{figure}
    \centering
    \includegraphics[width=0.95\columnwidth]{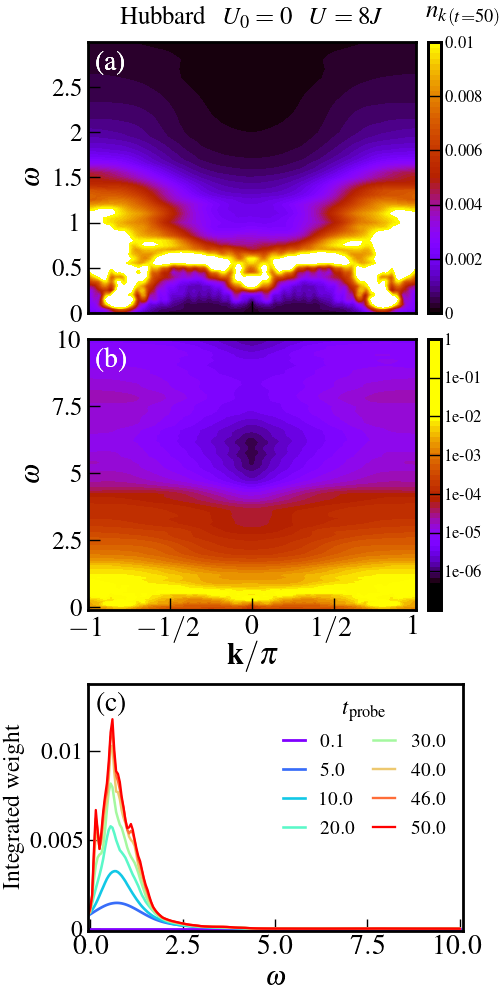}
    \caption{Neutron scattering of Hubbard chains quenched from $U/J=0$ to $U/J=8$. Momentum-resolved spectrum in (a) linear and (b) logarithmic color scales. Integrated weight as a function of probing times is shown in figure c).}
    \label{fig:spec_Hubbard_quench_U=8}
\end{figure}


\subsection{Hubbard chain after a quench}

We now proceed to studying the case of a Hubbard chain far from equilibrium, after a sudden quench in $H_{Hubbard}$ from $U/J=0$ to $U/J=8$. At $t=0$, the inital state is the ground-state of the non-interacting Hamiltonian. We then suddenly change the value of the interactions to $U/J=8$ and we measure the spectrum of the system in the resulting non-thermal state of the new interacting Hamiltonian. Much attention has been paid to the problem of the ``melting'' of the Mott insulator\cite{ Zawadzki2019,Kohno2010,Kohno2012, Nocera2018,Yang2016,kidd08,Matsueda2005,Zemljic2008,Eckstein2013, Eckstein2014, Belzer2015, Golez2015,Bittner2018}, mostly in the context of the photoemission response. By pumping energy into the system, one can change the population of doublons and induce excitations into the upper Hubbard band. The effects of the quench are similar to photo-doping: the chain is no longer insulating, but will have a finite density of holes and double occupied sites that will differ from that in equilibrium (essentially the equivalent to particle-hole excitations in a Mott insulator). As a result, the chain will be gapless, both for the spin and the charge sectors. This will be reflected in the spectrum probed by neutron scattering, that now will display a superposition of coexisting spin excitations in the upper and Hubbard band, as shown in our results, Fig.\ref{fig:spec_Hubbard_quench_U=8}. Interestingly, the high energy ``bubble'' has also ``melted'', together with the Mott gap. Consequently, the magnetic order (or "quasi order" in 1D) has also been modified: the signatures of ``$2k_F$'' singularities are no longer well defined and we see a indications of gapless dispersive branches shifted away from $k=\pi$, as expected from a doped Mott insulator\cite{Raczkowski2015,Bittner2018}. In a non-equilibrium non-thermal state such as the one realized in a quench, the concepts of bands or dispersion are not well defined in the conventional sense. The measured spectrum contains contributions from all allowed transitions $\omega_{mn}=E_m-E_n$, and will typically appear as an incoherent continuum.


\section{Conclusions}

We have presented a numerical approach to calculate inelastic scattering spectra by directly simulating a scattering event using the time-dependent Schr\"odinger equation. Unlike conventional approaches that rely on evaluating Green's functions in the frequency or time domain, we directly obtain the spectral density through the probability of detecting an event after an incident particle is deflected from the sample. The method not only reproduces the energy and momentum resolved results from equilibrium Green's functions, but includes contributions to all orders, revealing hidden features that can potentially be observed experimentally. These higher order features correspond to transitions between excited states. Their visibility depends on the magnitude of the coupling between the incident particles and the sample (the so-called ``matrix elements') and the intensity of the beam. For weak interactions (smaller J') they will be rapidly suppressed, since the next correction enters with a $J'^4$ prefactor. In terms of practicality in the context of numerical calculations, a smaller $J'$ implies a broadening in the spectral features for the same $t_{probe}$, meaning that we need to increase the simulation time to achieve the same resolution. The noteworthy aspect of this method is that, by circumventing the direct explicit evaluation of matrix elements between excited states, the approach can be readily and seamlessly applied to non-equilibrium problems that would otherwise be out of reach for conventional numerical alternatives.   

\acknowledgements
We thank Fabian Essler and Igor Zaliznyak for illuminating discussions and carefully reading the manuscript. We acknowledge generous computational resources provided by Northeastern University's Discovery Cluster at the Massachusetts Green High Performance Computing Center (MGHPCC). KZ is supported by a Faculty of the Future fellowship of the Schlumberger Foundation. AEF and LY are supported by the U.S. Department of Energy, Office of Basic Energy Sciences under grant No. DE-SC0014407. KZ is partially supported by the same grant. 

\appendix
\onecolumngrid
\section{Perturbative analysis of the response functions}
\label{apx:PT}
\subsection{Local probe}
    For simplicity, we first describe the local case in which a particle hits the sample at position ``0'' and interacts locally with the electrons via a local Coulomb term (effectively describing an EELS event).
    The first contribution to the number of particles with energy $\omega_d$ in the detector can be calculated as
    \begin{align}
        \label{eq:n2t}
        \braket{n_d(t)} & = \int_0^t dt_1 \int_0^t dt_2 
        \langle 
        e^{i (H_0+H_d)t_1} V e^{-i(H_0+H_d)t_1} n_d e^{i (H_0+H_d)t_2} V e^{-i(H_0+H_d)t_2}
        \rangle
        .
    \end{align}
    
     The action of $V$ on the initial state is very simple. Assuming that the system is initially in equilibrium in the ground-state:
        \begin{align}
            \label{eq:Vpsi0}
            V \ket{\Psi(t=0)} = & 
            J' n_0 \ket{gs}\ket{n_s=0,n_d=1},
        \end{align}
    and Eq. \ref{eq:n2t} becomes
    \begin{align}
        \label{eq:n2t-pt2}
        \braket{n_d(t)} & = {J'}^2
         \int_0^t dt_1 \int_0^t dt_2 
          e^{i(E_0 + \omega_s)(t_1-t_2)}
        \bra{gs} \bra{0,1} n_0 e^{-i(H_0+H_d)t_1} n_d e^{i(H_0+H_d)t_2}  n_0  \ket{gs} \ket{0,1}
        \nonumber \\
        & =  {J'}^2 \int_0^t dt_1 \int_0^t dt_2 
        e^{i(E_0 + \omega_s)(t_1-t_2)} e^{-i\omega_d(t_1-t_2)} \bra{gs} n_0  e^{-iH_0(t_1-t_2)} n_0 \ket{gs}
        \nonumber \\ 
        = & {J'}^2
        \int_0^t dt_1 \int_0^t dt_2 
         \sum_{\ket{f}}
        e^{i(E_0 + \omega_s-\omega_d)(t_1-t_2)} 
        e^{-iE_f (t_1-t_2)} |\bra{gs} n_0 \ket{f}|^2
        \nonumber \\ 
        = & {J'}^2
        \int_0^t dt_1 \int_0^t dt_2 
         \sum_{\ket{f}}
        e^{i((E_0 -E_f) + (\omega_s-\omega_d))(t_1-t_2)} 
       |\bra{gs} n_0 \ket{f}|^2 \nonumber \\
       =&4{J'}^2\sum_{\ket{f}}\frac{\sin^2{([\omega-(E_0-E_f)]t/2)}}{(\omega-(E_0-E_f))^2}|\bra{gs} n_0 \ket{f}|^2
       ,
    \end{align}    
    where $\omega=\omega_s-\omega_d$. In the limit of $t\rightarrow \infty$ it can be written as:
    \begin{equation}
        \frac{\langle n(t)}{t} \rightarrow 2\pi  \sum_{\ket{f}}|\bra{gs} n_0 \ket{f}|^2 \delta(\omega-(E_0-E_f)).
    \end{equation}
    Note that in the case of neutron scattering, we can replace the term $n_0 (n_s+n_d)$ in $V$ by 
    $S_0^z(S_1^z + S_2^z) = 1/2 S_0^2$, so that the observable in the brakets of the last line of 
    Eq. \ref{eq:n2t-pt2} is given by $S_0^z$.    

\subsection{Extended probe: momentum resolution}

    We generalize the previous case to an extended probe with momentum resolution, and an arbitrary contact term in the potential with an operator $O$. The signal at the detector is now the momentum distribution function, that can be obtained as 
    \begin{align}
        \label{eq:n2kt}
        \braket{n_{2k}(t)} & = \int_0^t dt_1 \int_0^t dt_2 
        \langle 
        e^{i (H_0+H_d)t_1} V e^{-i(H_0+H_d)t_1} n_{2k} e^{i (H_0+H_d)t_2} V e^{-i(H_0+H_d)t_2}
        \rangle
        .
    \end{align}    
    
    We assume that at $t=0$ the system is in the equilibrium in the ground state; hence: 
    \begin{align}
        \label{eq:n2kt-pt2}
        \braket{n_{2k}(t)} & = \int_0^t dt_1 \int_0^t dt_2 
        e^{i(E_0 + \omega_s) (t_1-t_2)}
        \bra{gs}\bra{k_0,0} V e^{-i(H_0+H_d)t_1} n_{2k} e^{i (H_0+H_d)t_2} V 
       \ket{gs}\ket{k_0,0}. 
    \end{align}        

    Applying $V$ to $\Psi(t=0)$ yields
    \begin{align}
        \label{eq:Vkpsi0k}
         V 
       \ket{gs}\ket{k_0,0}
        = & 
        \frac{J'}{L} \sum_\ell O_\ell\ket{gs} \sum_{q,p}  e^{i(p-q)\ell} c_{2p}^\dagger c_{1q} \ket{k_0,0}
        \nonumber \\
        = &
        \frac{J'}{L} \sum_\ell  \sum_{p}e^{i(p-k_0)\ell}O_\ell\ket{gs} \ket{0,p}
        .
    \end{align}
    In addition, $n_{2k}$ projects the state onto one with well defined momentum:
    \begin{equation}
        n_{2k}V\ket{gs}\ket{k_0,0} = \frac{J'}{L} \sum_\ell  e^{i(k-k_0)\ell}O_\ell\ket{gs} \ket{0,k} = J'O_{k-k_0}\ket{gs} \ket{0,k}.
    \end{equation}
    
    With that into consideration, expression \ref{eq:n2kt-pt2} becomes 
    \begin{align}
        \label{eq:n2kt-pt3}
        \braket{n_{2m}(t)} & = 
         J'^2 \sum_f \int_0^t dt_1 \int_0^t dt_2 
        e^{i(E_0 -E_f + \omega_s-\omega_d) (t_1-t_2)} 
        |\bra{f} O_{k-k_0} \ket{gs}|^2.
    \end{align}  
    In this expression we recognize the momentum resolved spectral function for operator $O$, shifted by $k_0$
    \begin{equation}
              \braket{n_{2m}(t)} =4{J'}^2\sum_{\ket{f}}\frac{\sin^2{([\omega-(E_0-E_f)]t/2)}}{(\omega-(E_0-E_f))^2}|\bra{gs} O_{k-k_0} \ket{f}|^2
    \end{equation}
 
\twocolumngrid
\bibliography{neutrons}

\begin{thebibliography}{67}%
\makeatletter
\providecommand \@ifxundefined [1]{%
 \@ifx{#1\undefined}
}%
\providecommand \@ifnum [1]{%
 \ifnum #1\expandafter \@firstoftwo
 \else \expandafter \@secondoftwo
 \fi
}%
\providecommand \@ifx [1]{%
 \ifx #1\expandafter \@firstoftwo
 \else \expandafter \@secondoftwo
 \fi
}%
\providecommand \natexlab [1]{#1}%
\providecommand \enquote  [1]{``#1''}%
\providecommand \bibnamefont  [1]{#1}%
\providecommand \bibfnamefont [1]{#1}%
\providecommand \citenamefont [1]{#1}%
\providecommand \href@noop [0]{\@secondoftwo}%
\providecommand \href [0]{\begingroup \@sanitize@url \@href}%
\providecommand \@href[1]{\@@startlink{#1}\@@href}%
\providecommand \@@href[1]{\endgroup#1\@@endlink}%
\providecommand \@sanitize@url [0]{\catcode `\\12\catcode `\$12\catcode
  `\&12\catcode `\#12\catcode `\^12\catcode `\_12\catcode `\%12\relax}%
\providecommand \@@startlink[1]{}%
\providecommand \@@endlink[0]{}%
\providecommand \url  [0]{\begingroup\@sanitize@url \@url }%
\providecommand \@url [1]{\endgroup\@href {#1}{\urlprefix }}%
\providecommand \urlprefix  [0]{URL }%
\providecommand \Eprint [0]{\href }%
\providecommand \doibase [0]{http://dx.doi.org/}%
\providecommand \selectlanguage [0]{\@gobble}%
\providecommand \bibinfo  [0]{\@secondoftwo}%
\providecommand \bibfield  [0]{\@secondoftwo}%
\providecommand \translation [1]{[#1]}%
\providecommand \BibitemOpen [0]{}%
\providecommand \bibitemStop [0]{}%
\providecommand \bibitemNoStop [0]{.\EOS\space}%
\providecommand \EOS [0]{\spacefactor3000\relax}%
\providecommand \BibitemShut  [1]{\csname bibitem#1\endcsname}%
\let\auto@bib@innerbib\@empty
\bibitem [{\citenamefont {Garc\'{\i}a~de Abajo}(2010)}]{eels1}%
  \BibitemOpen
  \bibfield  {author} {\bibinfo {author} {\bibfnamefont {F.~J.}\ \bibnamefont
  {Garc\'{\i}a~de Abajo}},\ }\href {\doibase 10.1103/RevModPhys.82.209}
  {\bibfield  {journal} {\bibinfo  {journal} {Rev. Mod. Phys.}\ }\textbf
  {\bibinfo {volume} {82}},\ \bibinfo {pages} {209} (\bibinfo {year}
  {2010})}\BibitemShut {NoStop}%
\bibitem [{\citenamefont {Hofer}\ \emph {et~al.}(2016)\citenamefont {Hofer},
  \citenamefont {Schmidt}, \citenamefont {Grogger},\ and\ \citenamefont
  {Kothleitner}}]{eels2}%
  \BibitemOpen
  \bibfield  {author} {\bibinfo {author} {\bibfnamefont {F.}~\bibnamefont
  {Hofer}}, \bibinfo {author} {\bibfnamefont {F.~P.}\ \bibnamefont {Schmidt}},
  \bibinfo {author} {\bibfnamefont {W.}~\bibnamefont {Grogger}}, \ and\
  \bibinfo {author} {\bibfnamefont {G.}~\bibnamefont {Kothleitner}},\ }\href
  {\doibase 10.1088/1757-899x/109/1/012007} {\bibfield  {journal} {\bibinfo
  {journal} {{IOP} Conference Series: Materials Science and Engineering}\
  }\textbf {\bibinfo {volume} {109}},\ \bibinfo {pages} {012007} (\bibinfo
  {year} {2016})}\BibitemShut {NoStop}%
\bibitem [{\citenamefont {Ritchie}\ and\ \citenamefont {Howie}(1988)}]{eels3}%
  \BibitemOpen
  \bibfield  {author} {\bibinfo {author} {\bibfnamefont {R.~H.}\ \bibnamefont
  {Ritchie}}\ and\ \bibinfo {author} {\bibfnamefont {A.}~\bibnamefont
  {Howie}},\ }\href {\doibase 10.1080/01418618808209951} {\bibfield  {journal}
  {\bibinfo  {journal} {Philosophical Magazine A}\ }\textbf {\bibinfo {volume}
  {58}},\ \bibinfo {pages} {753} (\bibinfo {year} {1988})},\ \Eprint
  {http://arxiv.org/abs/https://doi.org/10.1080/01418618808209951}
  {https://doi.org/10.1080/01418618808209951} \BibitemShut {NoStop}%
\bibitem [{\citenamefont {Egerton}(2011)}]{eels4}%
  \BibitemOpen
  \bibfield  {author} {\bibinfo {author} {\bibfnamefont {R.~F.}\ \bibnamefont
  {Egerton}},\ }\href@noop {} {\emph {\bibinfo {title} {Electron energy-loss
  spectroscopy in the electron microscope}}}\ (\bibinfo  {publisher}
  {Springer},\ \bibinfo {address} {Berlin, New York},\ \bibinfo {year}
  {2011})\BibitemShut {NoStop}%
\bibitem [{\citenamefont {Lippmann}\ and\ \citenamefont
  {Schwinger}(1950)}]{Lippmann1950}%
  \BibitemOpen
  \bibfield  {author} {\bibinfo {author} {\bibfnamefont {B.~A.}\ \bibnamefont
  {Lippmann}}\ and\ \bibinfo {author} {\bibfnamefont {J.}~\bibnamefont
  {Schwinger}},\ }\href {\doibase 10.1103/PhysRev.79.469} {\bibfield  {journal}
  {\bibinfo  {journal} {Phys. Rev.}\ }\textbf {\bibinfo {volume} {79}},\
  \bibinfo {pages} {469} (\bibinfo {year} {1950})}\BibitemShut {NoStop}%
\bibitem [{\citenamefont {Dagotto}(1994)}]{Dagotto1994}%
  \BibitemOpen
  \bibfield  {author} {\bibinfo {author} {\bibfnamefont {E.}~\bibnamefont
  {Dagotto}},\ }\href {\doibase 10.1103/RevModPhys.66.763} {\bibfield
  {journal} {\bibinfo  {journal} {Rev. Mod. Phys.}\ }\textbf {\bibinfo {volume}
  {66}},\ \bibinfo {pages} {763} (\bibinfo {year} {1994})}\BibitemShut
  {NoStop}%
\bibitem [{\citenamefont {Sch\"uttler}\ and\ \citenamefont
  {Scalapino}(1986)}]{Schuttler1986}%
  \BibitemOpen
  \bibfield  {author} {\bibinfo {author} {\bibfnamefont {H.~B.}\ \bibnamefont
  {Sch\"uttler}}\ and\ \bibinfo {author} {\bibfnamefont {D.~J.}\ \bibnamefont
  {Scalapino}},\ }\href {\doibase 10.1103/PhysRevB.34.4744} {\bibfield
  {journal} {\bibinfo  {journal} {Phys. Rev. B}\ }\textbf {\bibinfo {volume}
  {34}},\ \bibinfo {pages} {4744} (\bibinfo {year} {1986})}\BibitemShut
  {NoStop}%
\bibitem [{\citenamefont {Sandvik}(1998)}]{Sandvik1998}%
  \BibitemOpen
  \bibfield  {author} {\bibinfo {author} {\bibfnamefont {A.~W.}\ \bibnamefont
  {Sandvik}},\ }\href {\doibase 10.1103/PhysRevB.57.10287} {\bibfield
  {journal} {\bibinfo  {journal} {Phys. Rev. B}\ }\textbf {\bibinfo {volume}
  {57}},\ \bibinfo {pages} {10287} (\bibinfo {year} {1998})}\BibitemShut
  {NoStop}%
\bibitem [{\citenamefont {Silver}\ \emph {et~al.}(1990)\citenamefont {Silver},
  \citenamefont {Sivia},\ and\ \citenamefont {Gubernatis}}]{Silver1990}%
  \BibitemOpen
  \bibfield  {author} {\bibinfo {author} {\bibfnamefont {R.~N.}\ \bibnamefont
  {Silver}}, \bibinfo {author} {\bibfnamefont {D.~S.}\ \bibnamefont {Sivia}}, \
  and\ \bibinfo {author} {\bibfnamefont {J.~E.}\ \bibnamefont {Gubernatis}},\
  }\href {\doibase 10.1103/PhysRevB.41.2380} {\bibfield  {journal} {\bibinfo
  {journal} {Phys. Rev. B}\ }\textbf {\bibinfo {volume} {41}},\ \bibinfo
  {pages} {2380} (\bibinfo {year} {1990})}\BibitemShut {NoStop}%
\bibitem [{\citenamefont {Gubernatis}\ \emph {et~al.}(1991)\citenamefont
  {Gubernatis}, \citenamefont {Jarrell}, \citenamefont {Silver},\ and\
  \citenamefont {Sivia}}]{Gubernatis1991}%
  \BibitemOpen
  \bibfield  {author} {\bibinfo {author} {\bibfnamefont {J.~E.}\ \bibnamefont
  {Gubernatis}}, \bibinfo {author} {\bibfnamefont {M.}~\bibnamefont {Jarrell}},
  \bibinfo {author} {\bibfnamefont {R.~N.}\ \bibnamefont {Silver}}, \ and\
  \bibinfo {author} {\bibfnamefont {D.~S.}\ \bibnamefont {Sivia}},\ }\href
  {\doibase 10.1103/PhysRevB.44.6011} {\bibfield  {journal} {\bibinfo
  {journal} {Phys. Rev. B}\ }\textbf {\bibinfo {volume} {44}},\ \bibinfo
  {pages} {6011} (\bibinfo {year} {1991})}\BibitemShut {NoStop}%
\bibitem [{\citenamefont {Sylju\aa{}sen}(2008)}]{Syljuaasen2008}%
  \BibitemOpen
  \bibfield  {author} {\bibinfo {author} {\bibfnamefont {O.~F.}\ \bibnamefont
  {Sylju\aa{}sen}},\ }\href {\doibase 10.1103/PhysRevB.78.174429} {\bibfield
  {journal} {\bibinfo  {journal} {Phys. Rev. B}\ }\textbf {\bibinfo {volume}
  {78}},\ \bibinfo {pages} {174429} (\bibinfo {year} {2008})}\BibitemShut
  {NoStop}%
\bibitem [{\citenamefont {Fuchs}\ \emph {et~al.}(2010)\citenamefont {Fuchs},
  \citenamefont {Pruschke},\ and\ \citenamefont {Jarrell}}]{Fuchs2010}%
  \BibitemOpen
  \bibfield  {author} {\bibinfo {author} {\bibfnamefont {S.}~\bibnamefont
  {Fuchs}}, \bibinfo {author} {\bibfnamefont {T.}~\bibnamefont {Pruschke}}, \
  and\ \bibinfo {author} {\bibfnamefont {M.}~\bibnamefont {Jarrell}},\ }\href
  {\doibase 10.1103/PhysRevE.81.056701} {\bibfield  {journal} {\bibinfo
  {journal} {Phys. Rev. E}\ }\textbf {\bibinfo {volume} {81}},\ \bibinfo
  {pages} {056701} (\bibinfo {year} {2010})}\BibitemShut {NoStop}%
\bibitem [{\citenamefont {Sandvik}(2016)}]{Sandvik2016}%
  \BibitemOpen
  \bibfield  {author} {\bibinfo {author} {\bibfnamefont {A.~W.}\ \bibnamefont
  {Sandvik}},\ }\href {\doibase 10.1103/PhysRevE.94.063308} {\bibfield
  {journal} {\bibinfo  {journal} {Phys. Rev. E}\ }\textbf {\bibinfo {volume}
  {94}},\ \bibinfo {pages} {063308} (\bibinfo {year} {2016})}\BibitemShut
  {NoStop}%
\bibitem [{\citenamefont {Shao}\ \emph {et~al.}(2017)\citenamefont {Shao},
  \citenamefont {Qin}, \citenamefont {Capponi}, \citenamefont {Chesi},
  \citenamefont {Meng},\ and\ \citenamefont {Sandvik}}]{Shao2017}%
  \BibitemOpen
  \bibfield  {author} {\bibinfo {author} {\bibfnamefont {H.}~\bibnamefont
  {Shao}}, \bibinfo {author} {\bibfnamefont {Y.~Q.}\ \bibnamefont {Qin}},
  \bibinfo {author} {\bibfnamefont {S.}~\bibnamefont {Capponi}}, \bibinfo
  {author} {\bibfnamefont {S.}~\bibnamefont {Chesi}}, \bibinfo {author}
  {\bibfnamefont {Z.~Y.}\ \bibnamefont {Meng}}, \ and\ \bibinfo {author}
  {\bibfnamefont {A.~W.}\ \bibnamefont {Sandvik}},\ }\href {\doibase
  10.1103/PhysRevX.7.041072} {\bibfield  {journal} {\bibinfo  {journal} {Phys.
  Rev. X}\ }\textbf {\bibinfo {volume} {7}},\ \bibinfo {pages} {041072}
  (\bibinfo {year} {2017})}\BibitemShut {NoStop}%
\bibitem [{\citenamefont {Hallberg}(1995)}]{Hallberg1995}%
  \BibitemOpen
  \bibfield  {author} {\bibinfo {author} {\bibfnamefont {K.}~\bibnamefont
  {Hallberg}},\ }\href@noop {} {\bibfield  {journal} {\bibinfo  {journal}
  {Phys. Rev. B}\ }\textbf {\bibinfo {volume} {52}},\ \bibinfo {pages} {R9827}
  (\bibinfo {year} {1995})}\BibitemShut {NoStop}%
\bibitem [{\citenamefont {K\"uhner}\ and\ \citenamefont
  {White}(1999)}]{Kuhner1999}%
  \BibitemOpen
  \bibfield  {author} {\bibinfo {author} {\bibfnamefont {T.~D.}\ \bibnamefont
  {K\"uhner}}\ and\ \bibinfo {author} {\bibfnamefont {S.~R.}\ \bibnamefont
  {White}},\ }\href@noop {} {\bibfield  {journal} {\bibinfo  {journal} {Phys.
  Rev. B}\ }\textbf {\bibinfo {volume} {60}},\ \bibinfo {pages} {335} (\bibinfo
  {year} {1999})}\BibitemShut {NoStop}%
\bibitem [{\citenamefont {Jeckelmann}(2002)}]{Jeckelmann2002}%
  \BibitemOpen
  \bibfield  {author} {\bibinfo {author} {\bibfnamefont {E.}~\bibnamefont
  {Jeckelmann}},\ }\href@noop {} {\bibfield  {journal} {\bibinfo  {journal}
  {Phys. Rev. B}\ }\textbf {\bibinfo {volume} {66}},\ \bibinfo {pages} {045114}
  (\bibinfo {year} {2002})}\BibitemShut {NoStop}%
\bibitem [{\citenamefont {Daley}\ \emph {et~al.}(2004)\citenamefont {Daley},
  \citenamefont {Kollath}, \citenamefont {Schollw\"ock}, ,\ and\ \citenamefont
  {Vidal}}]{Daley2004}%
  \BibitemOpen
  \bibfield  {author} {\bibinfo {author} {\bibfnamefont {A.~J.}\ \bibnamefont
  {Daley}}, \bibinfo {author} {\bibfnamefont {C.}~\bibnamefont {Kollath}},
  \bibinfo {author} {\bibfnamefont {U.}~\bibnamefont {Schollw\"ock}}, , \ and\
  \bibinfo {author} {\bibfnamefont {G.}~\bibnamefont {Vidal}},\ }\href@noop {}
  {\bibfield  {journal} {\bibinfo  {journal} {J. Stat. Mech.: Theor. Exp.}\ ,\
  \bibinfo {pages} {P04005}} (\bibinfo {year} {2004})}\BibitemShut {NoStop}%
\bibitem [{\citenamefont {White}\ and\ \citenamefont
  {Feiguin}(2004)}]{White2004a}%
  \BibitemOpen
  \bibfield  {author} {\bibinfo {author} {\bibfnamefont {S.}~\bibnamefont
  {White}}\ and\ \bibinfo {author} {\bibfnamefont {A.}~\bibnamefont
  {Feiguin}},\ }\href@noop {} {\bibfield  {journal} {\bibinfo  {journal} {Phys.
  Rev. Lett.}\ }\textbf {\bibinfo {volume} {93}},\ \bibinfo {pages} {076401}
  (\bibinfo {year} {2004})}\BibitemShut {NoStop}%
\bibitem [{\citenamefont {Feiguin}\ and\ \citenamefont
  {White}(2005)}]{Feiguin2005}%
  \BibitemOpen
  \bibfield  {author} {\bibinfo {author} {\bibfnamefont {A.}~\bibnamefont
  {Feiguin}}\ and\ \bibinfo {author} {\bibfnamefont {S.}~\bibnamefont
  {White}},\ }\href@noop {} {\bibfield  {journal} {\bibinfo  {journal} {Phys.
  Rev. B}\ }\textbf {\bibinfo {volume} {72}},\ \bibinfo {pages} {20404}
  (\bibinfo {year} {2005})}\BibitemShut {NoStop}%
\bibitem [{\citenamefont {Feiguin}(2011)}]{vietri}%
  \BibitemOpen
  \bibfield  {author} {\bibinfo {author} {\bibfnamefont {A.~E.}\ \bibnamefont
  {Feiguin}},\ }in\ \href@noop {} {\emph {\bibinfo {booktitle} {XV Training
  Course in the Physics of Strongly Correlated Systems}}},\ Vol.\ \bibinfo
  {volume} {1419}\ (\bibinfo  {publisher} {AIP Proceedings},\ \bibinfo {year}
  {2011})\ p.~\bibinfo {pages} {5}\BibitemShut {NoStop}%
\bibitem [{\citenamefont {Feiguin}(2013)}]{Feiguin2013b}%
  \BibitemOpen
  \bibfield  {author} {\bibinfo {author} {\bibfnamefont {A.~E.}\ \bibnamefont
  {Feiguin}},\ }in\ \href@noop {} {\emph {\bibinfo {booktitle} {Strongly
  correlated systems: Numerical methods}}},\ \bibinfo {editor} {edited by\
  \bibinfo {editor} {\bibfnamefont {A.}~\bibnamefont {Avella}}\ and\ \bibinfo
  {editor} {\bibfnamefont {F.}~\bibnamefont {Mancini}}}\ (\bibinfo  {publisher}
  {Springer, Heidelberg, Berlin},\ \bibinfo {year} {2013})\BibitemShut
  {NoStop}%
\bibitem [{\citenamefont {Paeckel}\ \emph {et~al.}(2019)\citenamefont
  {Paeckel}, \citenamefont {Köhler}, \citenamefont {Swoboda}, \citenamefont
  {Manmana}, \citenamefont {Schollwöck},\ and\ \citenamefont
  {Hubig}}]{PAECKEL2019}%
  \BibitemOpen
  \bibfield  {author} {\bibinfo {author} {\bibfnamefont {S.}~\bibnamefont
  {Paeckel}}, \bibinfo {author} {\bibfnamefont {T.}~\bibnamefont {Köhler}},
  \bibinfo {author} {\bibfnamefont {A.}~\bibnamefont {Swoboda}}, \bibinfo
  {author} {\bibfnamefont {S.~R.}\ \bibnamefont {Manmana}}, \bibinfo {author}
  {\bibfnamefont {U.}~\bibnamefont {Schollwöck}}, \ and\ \bibinfo {author}
  {\bibfnamefont {C.}~\bibnamefont {Hubig}},\ }\href {\doibase
  https://doi.org/10.1016/j.aop.2019.167998} {\bibfield  {journal} {\bibinfo
  {journal} {Annals of Physics}\ }\textbf {\bibinfo {volume} {411}},\ \bibinfo
  {pages} {167998} (\bibinfo {year} {2019})}\BibitemShut {NoStop}%
\bibitem [{\citenamefont {Holzner}\ \emph {et~al.}(2011)\citenamefont
  {Holzner}, \citenamefont {Weichselbaum}, \citenamefont {McCulloch},
  \citenamefont {Schollw\"ock},\ and\ \citenamefont {von Delft}}]{Holzner2011}%
  \BibitemOpen
  \bibfield  {author} {\bibinfo {author} {\bibfnamefont {A.}~\bibnamefont
  {Holzner}}, \bibinfo {author} {\bibfnamefont {A.}~\bibnamefont
  {Weichselbaum}}, \bibinfo {author} {\bibfnamefont {I.~P.}\ \bibnamefont
  {McCulloch}}, \bibinfo {author} {\bibfnamefont {U.}~\bibnamefont
  {Schollw\"ock}}, \ and\ \bibinfo {author} {\bibfnamefont {J.}~\bibnamefont
  {von Delft}},\ }\href {\doibase 10.1103/PhysRevB.83.195115} {\bibfield
  {journal} {\bibinfo  {journal} {Phys. Rev. B}\ }\textbf {\bibinfo {volume}
  {83}},\ \bibinfo {pages} {195115} (\bibinfo {year} {2011})}\BibitemShut
  {NoStop}%
\bibitem [{\citenamefont {Wolf}\ \emph {et~al.}(2015)\citenamefont {Wolf},
  \citenamefont {Justiniano}, \citenamefont {McCulloch},\ and\ \citenamefont
  {Schollw\"ock}}]{Wolf2015}%
  \BibitemOpen
  \bibfield  {author} {\bibinfo {author} {\bibfnamefont {F.~A.}\ \bibnamefont
  {Wolf}}, \bibinfo {author} {\bibfnamefont {J.~A.}\ \bibnamefont
  {Justiniano}}, \bibinfo {author} {\bibfnamefont {I.~P.}\ \bibnamefont
  {McCulloch}}, \ and\ \bibinfo {author} {\bibfnamefont {U.}~\bibnamefont
  {Schollw\"ock}},\ }\href {\doibase 10.1103/PhysRevB.91.115144} {\bibfield
  {journal} {\bibinfo  {journal} {Phys. Rev. B}\ }\textbf {\bibinfo {volume}
  {91}},\ \bibinfo {pages} {115144} (\bibinfo {year} {2015})}\BibitemShut
  {NoStop}%
\bibitem [{\citenamefont {Xie}\ \emph {et~al.}(2018)\citenamefont {Xie},
  \citenamefont {Huang}, \citenamefont {Han}, \citenamefont {Yan},
  \citenamefont {Zhao}, \citenamefont {Xie}, \citenamefont {Liao},\ and\
  \citenamefont {Xiang}}]{Xie2018}%
  \BibitemOpen
  \bibfield  {author} {\bibinfo {author} {\bibfnamefont {H.~D.}\ \bibnamefont
  {Xie}}, \bibinfo {author} {\bibfnamefont {R.~Z.}\ \bibnamefont {Huang}},
  \bibinfo {author} {\bibfnamefont {X.~J.}\ \bibnamefont {Han}}, \bibinfo
  {author} {\bibfnamefont {X.}~\bibnamefont {Yan}}, \bibinfo {author}
  {\bibfnamefont {H.~H.}\ \bibnamefont {Zhao}}, \bibinfo {author}
  {\bibfnamefont {Z.~Y.}\ \bibnamefont {Xie}}, \bibinfo {author} {\bibfnamefont
  {H.~J.}\ \bibnamefont {Liao}}, \ and\ \bibinfo {author} {\bibfnamefont
  {T.}~\bibnamefont {Xiang}},\ }\href {\doibase 10.1103/PhysRevB.97.075111}
  {\bibfield  {journal} {\bibinfo  {journal} {Phys. Rev. B}\ }\textbf {\bibinfo
  {volume} {97}},\ \bibinfo {pages} {075111} (\bibinfo {year}
  {2018})}\BibitemShut {NoStop}%
\bibitem [{\citenamefont {Vanderstraeten}\ \emph
  {et~al.}(2015{\natexlab{a}})\citenamefont {Vanderstraeten}, \citenamefont
  {Mari\"en}, \citenamefont {Verstraete},\ and\ \citenamefont
  {Haegeman}}]{Vanderstraeten2015}%
  \BibitemOpen
  \bibfield  {author} {\bibinfo {author} {\bibfnamefont {L.}~\bibnamefont
  {Vanderstraeten}}, \bibinfo {author} {\bibfnamefont {M.}~\bibnamefont
  {Mari\"en}}, \bibinfo {author} {\bibfnamefont {F.}~\bibnamefont
  {Verstraete}}, \ and\ \bibinfo {author} {\bibfnamefont {J.}~\bibnamefont
  {Haegeman}},\ }\href {\doibase 10.1103/PhysRevB.92.201111} {\bibfield
  {journal} {\bibinfo  {journal} {Phys. Rev. B}\ }\textbf {\bibinfo {volume}
  {92}},\ \bibinfo {pages} {201111} (\bibinfo {year}
  {2015}{\natexlab{a}})}\BibitemShut {NoStop}%
\bibitem [{\citenamefont {Vanderstraeten}\ \emph
  {et~al.}(2015{\natexlab{b}})\citenamefont {Vanderstraeten}, \citenamefont
  {Verstraete},\ and\ \citenamefont {Haegeman}}]{Vanderstraeten2015b}%
  \BibitemOpen
  \bibfield  {author} {\bibinfo {author} {\bibfnamefont {L.}~\bibnamefont
  {Vanderstraeten}}, \bibinfo {author} {\bibfnamefont {F.}~\bibnamefont
  {Verstraete}}, \ and\ \bibinfo {author} {\bibfnamefont {J.}~\bibnamefont
  {Haegeman}},\ }\href {\doibase 10.1103/PhysRevB.92.125136} {\bibfield
  {journal} {\bibinfo  {journal} {Phys. Rev. B}\ }\textbf {\bibinfo {volume}
  {92}},\ \bibinfo {pages} {125136} (\bibinfo {year}
  {2015}{\natexlab{b}})}\BibitemShut {NoStop}%
\bibitem [{\citenamefont {Li}\ and\ \citenamefont {Yang}(2010)}]{Li2010}%
  \BibitemOpen
  \bibfield  {author} {\bibinfo {author} {\bibfnamefont {T.}~\bibnamefont
  {Li}}\ and\ \bibinfo {author} {\bibfnamefont {F.}~\bibnamefont {Yang}},\
  }\href {\doibase 10.1103/PhysRevB.81.214509} {\bibfield  {journal} {\bibinfo
  {journal} {Phys. Rev. B}\ }\textbf {\bibinfo {volume} {81}},\ \bibinfo
  {pages} {214509} (\bibinfo {year} {2010})}\BibitemShut {NoStop}%
\bibitem [{\citenamefont {Dalla~Piazza}\ \emph {et~al.}(2014)\citenamefont
  {Dalla~Piazza}, \citenamefont {Mourigal}, \citenamefont {Christensen},
  \citenamefont {Nilsen}, \citenamefont {Tregenna-Piggott}, \citenamefont
  {Perring}, \citenamefont {Enderle}, \citenamefont {McMorrow}, \citenamefont
  {Ivanov},\ and\ \citenamefont {R{\o}nnow}}]{DallaPiazza2014}%
  \BibitemOpen
  \bibfield  {author} {\bibinfo {author} {\bibfnamefont {B.}~\bibnamefont
  {Dalla~Piazza}}, \bibinfo {author} {\bibfnamefont {M.}~\bibnamefont
  {Mourigal}}, \bibinfo {author} {\bibfnamefont {N.~B.}\ \bibnamefont
  {Christensen}}, \bibinfo {author} {\bibfnamefont {G.~J.}\ \bibnamefont
  {Nilsen}}, \bibinfo {author} {\bibfnamefont {P.}~\bibnamefont
  {Tregenna-Piggott}}, \bibinfo {author} {\bibfnamefont {T.~G.}\ \bibnamefont
  {Perring}}, \bibinfo {author} {\bibfnamefont {M.}~\bibnamefont {Enderle}},
  \bibinfo {author} {\bibfnamefont {D.~F.}\ \bibnamefont {McMorrow}}, \bibinfo
  {author} {\bibfnamefont {D.~A.}\ \bibnamefont {Ivanov}}, \ and\ \bibinfo
  {author} {\bibfnamefont {H.~M.}\ \bibnamefont {R{\o}nnow}},\ }\href
  {https://doi.org/10.1038/nphys3172} {\bibfield  {journal} {\bibinfo
  {journal} {Nature Physics}\ }\textbf {\bibinfo {volume} {11}},\ \bibinfo
  {pages} {62 EP } (\bibinfo {year} {2014})},\ \bibinfo {note}
  {article}\BibitemShut {NoStop}%
\bibitem [{\citenamefont {Ferrari}\ \emph {et~al.}(2018)\citenamefont
  {Ferrari}, \citenamefont {Parola}, \citenamefont {Sorella},\ and\
  \citenamefont {Becca}}]{Ferrari2018}%
  \BibitemOpen
  \bibfield  {author} {\bibinfo {author} {\bibfnamefont {F.}~\bibnamefont
  {Ferrari}}, \bibinfo {author} {\bibfnamefont {A.}~\bibnamefont {Parola}},
  \bibinfo {author} {\bibfnamefont {S.}~\bibnamefont {Sorella}}, \ and\
  \bibinfo {author} {\bibfnamefont {F.}~\bibnamefont {Becca}},\ }\href
  {\doibase 10.1103/PhysRevB.97.235103} {\bibfield  {journal} {\bibinfo
  {journal} {Phys. Rev. B}\ }\textbf {\bibinfo {volume} {97}},\ \bibinfo
  {pages} {235103} (\bibinfo {year} {2018})}\BibitemShut {NoStop}%
\bibitem [{\citenamefont {Hendry}\ and\ \citenamefont
  {Feiguin}(2019)}]{Hendry2020}%
  \BibitemOpen
  \bibfield  {author} {\bibinfo {author} {\bibfnamefont {D.}~\bibnamefont
  {Hendry}}\ and\ \bibinfo {author} {\bibfnamefont {A.~E.}\ \bibnamefont
  {Feiguin}},\ }\href {\doibase 10.1103/PhysRevB.100.245123} {\bibfield
  {journal} {\bibinfo  {journal} {Phys. Rev. B}\ }\textbf {\bibinfo {volume}
  {100}},\ \bibinfo {pages} {245123} (\bibinfo {year} {2019})}\BibitemShut
  {NoStop}%
\bibitem [{\citenamefont {Zawadzki}\ and\ \citenamefont
  {Feiguin}(2019)}]{Zawadzki2019}%
  \BibitemOpen
  \bibfield  {author} {\bibinfo {author} {\bibfnamefont {K.}~\bibnamefont
  {Zawadzki}}\ and\ \bibinfo {author} {\bibfnamefont {A.~E.}\ \bibnamefont
  {Feiguin}},\ }\href {\doibase 10.1103/PhysRevB.100.195124} {\bibfield
  {journal} {\bibinfo  {journal} {Phys. Rev. B}\ }\textbf {\bibinfo {volume}
  {100}},\ \bibinfo {pages} {195124} (\bibinfo {year} {2019})}\BibitemShut
  {NoStop}%
\bibitem [{\citenamefont {Freericks}\ \emph {et~al.}(2009)\citenamefont
  {Freericks}, \citenamefont {Krishnamurthy},\ and\ \citenamefont
  {Pruschke}}]{Freericks2009}%
  \BibitemOpen
  \bibfield  {author} {\bibinfo {author} {\bibfnamefont {J.~K.}\ \bibnamefont
  {Freericks}}, \bibinfo {author} {\bibfnamefont {H.~R.}\ \bibnamefont
  {Krishnamurthy}}, \ and\ \bibinfo {author} {\bibfnamefont {T.}~\bibnamefont
  {Pruschke}},\ }\href {\doibase 10.1103/PhysRevLett.102.136401} {\bibfield
  {journal} {\bibinfo  {journal} {Phys. Rev. Lett.}\ }\textbf {\bibinfo
  {volume} {102}},\ \bibinfo {pages} {136401} (\bibinfo {year}
  {2009})}\BibitemShut {NoStop}%
\bibitem [{\citenamefont {Shao}\ \emph {et~al.}(2016)\citenamefont {Shao},
  \citenamefont {Tohyama}, \citenamefont {Luo},\ and\ \citenamefont
  {Lu}}]{Shao2016}%
  \BibitemOpen
  \bibfield  {author} {\bibinfo {author} {\bibfnamefont {C.}~\bibnamefont
  {Shao}}, \bibinfo {author} {\bibfnamefont {T.}~\bibnamefont {Tohyama}},
  \bibinfo {author} {\bibfnamefont {H.-G.}\ \bibnamefont {Luo}}, \ and\
  \bibinfo {author} {\bibfnamefont {H.}~\bibnamefont {Lu}},\ }\href {\doibase
  10.1103/PhysRevB.93.195144} {\bibfield  {journal} {\bibinfo  {journal} {Phys.
  Rev. B}\ }\textbf {\bibinfo {volume} {93}},\ \bibinfo {pages} {195144}
  (\bibinfo {year} {2016})}\BibitemShut {NoStop}%
\bibitem [{\citenamefont {Zawadzki}\ \emph {et~al.}(2020)\citenamefont
  {Zawadzki}, \citenamefont {Nocera},\ and\ \citenamefont
  {Feiguin}}]{Zawadzki2020}%
  \BibitemOpen
  \bibfield  {author} {\bibinfo {author} {\bibfnamefont {K.}~\bibnamefont
  {Zawadzki}}, \bibinfo {author} {\bibfnamefont {A.}~\bibnamefont {Nocera}}, \
  and\ \bibinfo {author} {\bibfnamefont {A.~E.}\ \bibnamefont {Feiguin}},\
  }\href@noop {} {\enquote {\bibinfo {title} {A time-dependent scattering
  approach to core-level spectroscopies},}\ } (\bibinfo {year} {2020}),\
  \Eprint {http://arxiv.org/abs/2002.04142} {arXiv:2002.04142
  [cond-mat.str-el]} \BibitemShut {NoStop}%
\bibitem [{\citenamefont {des Cloizeaux}\ and\ \citenamefont
  {Pearson}(1962)}]{Cloizeaux1962}%
  \BibitemOpen
  \bibfield  {author} {\bibinfo {author} {\bibfnamefont {J.}~\bibnamefont {des
  Cloizeaux}}\ and\ \bibinfo {author} {\bibfnamefont {J.~J.}\ \bibnamefont
  {Pearson}},\ }\href {\doibase 10.1103/PhysRev.128.2131} {\bibfield  {journal}
  {\bibinfo  {journal} {Phys. Rev.}\ }\textbf {\bibinfo {volume} {128}},\
  \bibinfo {pages} {2131} (\bibinfo {year} {1962})}\BibitemShut {NoStop}%
\bibitem [{\citenamefont {M\"uller}\ \emph {et~al.}(1981)\citenamefont
  {M\"uller}, \citenamefont {Thomas}, \citenamefont {Beck},\ and\ \citenamefont
  {Bonner}}]{Muller1981}%
  \BibitemOpen
  \bibfield  {author} {\bibinfo {author} {\bibfnamefont {G.}~\bibnamefont
  {M\"uller}}, \bibinfo {author} {\bibfnamefont {H.}~\bibnamefont {Thomas}},
  \bibinfo {author} {\bibfnamefont {H.}~\bibnamefont {Beck}}, \ and\ \bibinfo
  {author} {\bibfnamefont {J.~C.}\ \bibnamefont {Bonner}},\ }\href {\doibase
  10.1103/PhysRevB.24.1429} {\bibfield  {journal} {\bibinfo  {journal} {Phys.
  Rev. B}\ }\textbf {\bibinfo {volume} {24}},\ \bibinfo {pages} {1429}
  (\bibinfo {year} {1981})}\BibitemShut {NoStop}%
\bibitem [{\citenamefont {Zheludev}\ \emph {et~al.}(2001)\citenamefont
  {Zheludev}, \citenamefont {Kenzelmann}, \citenamefont {Raymond},
  \citenamefont {Masuda}, \citenamefont {Uchinokura},\ and\ \citenamefont
  {Lee}}]{Zheludev2001}%
  \BibitemOpen
  \bibfield  {author} {\bibinfo {author} {\bibfnamefont {A.}~\bibnamefont
  {Zheludev}}, \bibinfo {author} {\bibfnamefont {M.}~\bibnamefont
  {Kenzelmann}}, \bibinfo {author} {\bibfnamefont {S.}~\bibnamefont {Raymond}},
  \bibinfo {author} {\bibfnamefont {T.}~\bibnamefont {Masuda}}, \bibinfo
  {author} {\bibfnamefont {K.}~\bibnamefont {Uchinokura}}, \ and\ \bibinfo
  {author} {\bibfnamefont {S.-H.}\ \bibnamefont {Lee}},\ }\href {\doibase
  10.1103/PhysRevB.65.014402} {\bibfield  {journal} {\bibinfo  {journal} {Phys.
  Rev. B}\ }\textbf {\bibinfo {volume} {65}},\ \bibinfo {pages} {014402}
  (\bibinfo {year} {2001})}\BibitemShut {NoStop}%
\bibitem [{\citenamefont {Stone}\ \emph {et~al.}(2003)\citenamefont {Stone},
  \citenamefont {Reich}, \citenamefont {Broholm}, \citenamefont {Lefmann},
  \citenamefont {Rischel}, \citenamefont {Landee},\ and\ \citenamefont
  {Turnbull}}]{Stone2003}%
  \BibitemOpen
  \bibfield  {author} {\bibinfo {author} {\bibfnamefont {M.}~\bibnamefont
  {Stone}}, \bibinfo {author} {\bibfnamefont {D.}~\bibnamefont {Reich}},
  \bibinfo {author} {\bibfnamefont {C.}~\bibnamefont {Broholm}}, \bibinfo
  {author} {\bibfnamefont {K.}~\bibnamefont {Lefmann}}, \bibinfo {author}
  {\bibfnamefont {C.}~\bibnamefont {Rischel}}, \bibinfo {author} {\bibfnamefont
  {C.}~\bibnamefont {Landee}}, \ and\ \bibinfo {author} {\bibfnamefont
  {M.}~\bibnamefont {Turnbull}},\ }\href {\doibase
  10.1103/PhysRevLett.91.037205} {\bibfield  {journal} {\bibinfo  {journal}
  {Physical review letters}\ }\textbf {\bibinfo {volume} {91}},\ \bibinfo
  {pages} {037205} (\bibinfo {year} {2003})}\BibitemShut {NoStop}%
\bibitem [{\citenamefont {Kenzelmann}\ \emph {et~al.}(2004)\citenamefont
  {Kenzelmann}, \citenamefont {Chen}, \citenamefont {Broholm}, \citenamefont
  {Reich},\ and\ \citenamefont {Qiu}}]{Kenzelmann2004}%
  \BibitemOpen
  \bibfield  {author} {\bibinfo {author} {\bibfnamefont {M.}~\bibnamefont
  {Kenzelmann}}, \bibinfo {author} {\bibfnamefont {Y.}~\bibnamefont {Chen}},
  \bibinfo {author} {\bibfnamefont {C.}~\bibnamefont {Broholm}}, \bibinfo
  {author} {\bibfnamefont {D.~H.}\ \bibnamefont {Reich}}, \ and\ \bibinfo
  {author} {\bibfnamefont {Y.}~\bibnamefont {Qiu}},\ }\href {\doibase
  10.1103/PhysRevLett.93.017204} {\bibfield  {journal} {\bibinfo  {journal}
  {Phys. Rev. Lett.}\ }\textbf {\bibinfo {volume} {93}},\ \bibinfo {pages}
  {017204} (\bibinfo {year} {2004})}\BibitemShut {NoStop}%
\bibitem [{\citenamefont {Kohno}(2009)}]{Kohno2009}%
  \BibitemOpen
  \bibfield  {author} {\bibinfo {author} {\bibfnamefont {M.}~\bibnamefont
  {Kohno}},\ }\href {\doibase 10.1103/PhysRevLett.102.037203} {\bibfield
  {journal} {\bibinfo  {journal} {Phys. Rev. Lett.}\ }\textbf {\bibinfo
  {volume} {102}},\ \bibinfo {pages} {037203} (\bibinfo {year}
  {2009})}\BibitemShut {NoStop}%
\bibitem [{\citenamefont {Lake}\ \emph {et~al.}(2010)\citenamefont {Lake},
  \citenamefont {Tsvelik}, \citenamefont {Notbohm}, \citenamefont
  {Alan~Tennant}, \citenamefont {Perring}, \citenamefont {Reehuis},
  \citenamefont {Sekar}, \citenamefont {Krabbes},\ and\ \citenamefont
  {B{\"u}chner}}]{Lake2010}%
  \BibitemOpen
  \bibfield  {author} {\bibinfo {author} {\bibfnamefont {B.}~\bibnamefont
  {Lake}}, \bibinfo {author} {\bibfnamefont {A.~M.}\ \bibnamefont {Tsvelik}},
  \bibinfo {author} {\bibfnamefont {S.}~\bibnamefont {Notbohm}}, \bibinfo
  {author} {\bibfnamefont {D.}~\bibnamefont {Alan~Tennant}}, \bibinfo {author}
  {\bibfnamefont {T.~G.}\ \bibnamefont {Perring}}, \bibinfo {author}
  {\bibfnamefont {M.}~\bibnamefont {Reehuis}}, \bibinfo {author} {\bibfnamefont
  {C.}~\bibnamefont {Sekar}}, \bibinfo {author} {\bibfnamefont
  {G.}~\bibnamefont {Krabbes}}, \ and\ \bibinfo {author} {\bibfnamefont
  {B.}~\bibnamefont {B{\"u}chner}},\ }\href {\doibase 10.1038/nphys1462}
  {\bibfield  {journal} {\bibinfo  {journal} {Nature Physics}\ }\textbf
  {\bibinfo {volume} {6}},\ \bibinfo {pages} {50} (\bibinfo {year}
  {2010})}\BibitemShut {NoStop}%
\bibitem [{\citenamefont {Bouillot}\ \emph {et~al.}(2011)\citenamefont
  {Bouillot}, \citenamefont {Kollath}, \citenamefont {L\"auchli}, \citenamefont
  {Zvonarev}, \citenamefont {Thielemann}, \citenamefont {R\"uegg},
  \citenamefont {Orignac}, \citenamefont {Citro}, \citenamefont
  {Klanj\ifmmode~\check{s}\else \v{s}\fi{}ek}, \citenamefont {Berthier},
  \citenamefont {Horvati\ifmmode~\acute{c}\else \'{c}\fi{}},\ and\
  \citenamefont {Giamarchi}}]{Bouillot2011}%
  \BibitemOpen
  \bibfield  {author} {\bibinfo {author} {\bibfnamefont {P.}~\bibnamefont
  {Bouillot}}, \bibinfo {author} {\bibfnamefont {C.}~\bibnamefont {Kollath}},
  \bibinfo {author} {\bibfnamefont {A.~M.}\ \bibnamefont {L\"auchli}}, \bibinfo
  {author} {\bibfnamefont {M.}~\bibnamefont {Zvonarev}}, \bibinfo {author}
  {\bibfnamefont {B.}~\bibnamefont {Thielemann}}, \bibinfo {author}
  {\bibfnamefont {C.}~\bibnamefont {R\"uegg}}, \bibinfo {author} {\bibfnamefont
  {E.}~\bibnamefont {Orignac}}, \bibinfo {author} {\bibfnamefont
  {R.}~\bibnamefont {Citro}}, \bibinfo {author} {\bibfnamefont
  {M.}~\bibnamefont {Klanj\ifmmode~\check{s}\else \v{s}\fi{}ek}}, \bibinfo
  {author} {\bibfnamefont {C.}~\bibnamefont {Berthier}}, \bibinfo {author}
  {\bibfnamefont {M.}~\bibnamefont {Horvati\ifmmode~\acute{c}\else
  \'{c}\fi{}}}, \ and\ \bibinfo {author} {\bibfnamefont {T.}~\bibnamefont
  {Giamarchi}},\ }\href {\doibase 10.1103/PhysRevB.83.054407} {\bibfield
  {journal} {\bibinfo  {journal} {Phys. Rev. B}\ }\textbf {\bibinfo {volume}
  {83}},\ \bibinfo {pages} {054407} (\bibinfo {year} {2011})}\BibitemShut
  {NoStop}%
\bibitem [{\citenamefont {Schmidiger}\ \emph
  {et~al.}(2013{\natexlab{a}})\citenamefont {Schmidiger}, \citenamefont
  {Bouillot}, \citenamefont {Guidi}, \citenamefont {Bewley}, \citenamefont
  {Kollath}, \citenamefont {Giamarchi},\ and\ \citenamefont
  {Zheludev}}]{Schmidiger2013}%
  \BibitemOpen
  \bibfield  {author} {\bibinfo {author} {\bibfnamefont {D.}~\bibnamefont
  {Schmidiger}}, \bibinfo {author} {\bibfnamefont {P.}~\bibnamefont
  {Bouillot}}, \bibinfo {author} {\bibfnamefont {T.}~\bibnamefont {Guidi}},
  \bibinfo {author} {\bibfnamefont {R.}~\bibnamefont {Bewley}}, \bibinfo
  {author} {\bibfnamefont {C.}~\bibnamefont {Kollath}}, \bibinfo {author}
  {\bibfnamefont {T.}~\bibnamefont {Giamarchi}}, \ and\ \bibinfo {author}
  {\bibfnamefont {A.}~\bibnamefont {Zheludev}},\ }\href {\doibase
  10.1103/PhysRevLett.111.107202} {\bibfield  {journal} {\bibinfo  {journal}
  {Phys. Rev. Lett.}\ }\textbf {\bibinfo {volume} {111}},\ \bibinfo {pages}
  {107202} (\bibinfo {year} {2013}{\natexlab{a}})}\BibitemShut {NoStop}%
\bibitem [{\citenamefont {Schmidiger}\ \emph
  {et~al.}(2013{\natexlab{b}})\citenamefont {Schmidiger}, \citenamefont
  {M\"uhlbauer}, \citenamefont {Zheludev}, \citenamefont {Bouillot},
  \citenamefont {Giamarchi}, \citenamefont {Kollath}, \citenamefont {Ehlers},\
  and\ \citenamefont {Tsvelik}}]{Schmidiger2013b}%
  \BibitemOpen
  \bibfield  {author} {\bibinfo {author} {\bibfnamefont {D.}~\bibnamefont
  {Schmidiger}}, \bibinfo {author} {\bibfnamefont {S.}~\bibnamefont
  {M\"uhlbauer}}, \bibinfo {author} {\bibfnamefont {A.}~\bibnamefont
  {Zheludev}}, \bibinfo {author} {\bibfnamefont {P.}~\bibnamefont {Bouillot}},
  \bibinfo {author} {\bibfnamefont {T.}~\bibnamefont {Giamarchi}}, \bibinfo
  {author} {\bibfnamefont {C.}~\bibnamefont {Kollath}}, \bibinfo {author}
  {\bibfnamefont {G.}~\bibnamefont {Ehlers}}, \ and\ \bibinfo {author}
  {\bibfnamefont {A.~M.}\ \bibnamefont {Tsvelik}},\ }\href {\doibase
  10.1103/PhysRevB.88.094411} {\bibfield  {journal} {\bibinfo  {journal} {Phys.
  Rev. B}\ }\textbf {\bibinfo {volume} {88}},\ \bibinfo {pages} {094411}
  (\bibinfo {year} {2013}{\natexlab{b}})}\BibitemShut {NoStop}%
\bibitem [{\citenamefont {Casola}\ \emph {et~al.}(2013)\citenamefont {Casola},
  \citenamefont {Shiroka}, \citenamefont {Feiguin}, \citenamefont {Wang},
  \citenamefont {Grbi\ifmmode~\acute{c}\else \'{c}\fi{}}, \citenamefont
  {Horvati\ifmmode~\acute{c}\else \'{c}\fi{}}, \citenamefont {Kr\"amer},
  \citenamefont {Mukhopadhyay}, \citenamefont {Conder}, \citenamefont
  {Berthier}, \citenamefont {Ott}, \citenamefont {R\o{}nnow}, \citenamefont
  {R\"uegg},\ and\ \citenamefont {Mesot}}]{casola2013field}%
  \BibitemOpen
  \bibfield  {author} {\bibinfo {author} {\bibfnamefont {F.}~\bibnamefont
  {Casola}}, \bibinfo {author} {\bibfnamefont {T.}~\bibnamefont {Shiroka}},
  \bibinfo {author} {\bibfnamefont {A.}~\bibnamefont {Feiguin}}, \bibinfo
  {author} {\bibfnamefont {S.}~\bibnamefont {Wang}}, \bibinfo {author}
  {\bibfnamefont {M.~S.}\ \bibnamefont {Grbi\ifmmode~\acute{c}\else
  \'{c}\fi{}}}, \bibinfo {author} {\bibfnamefont {M.}~\bibnamefont
  {Horvati\ifmmode~\acute{c}\else \'{c}\fi{}}}, \bibinfo {author}
  {\bibfnamefont {S.}~\bibnamefont {Kr\"amer}}, \bibinfo {author}
  {\bibfnamefont {S.}~\bibnamefont {Mukhopadhyay}}, \bibinfo {author}
  {\bibfnamefont {K.}~\bibnamefont {Conder}}, \bibinfo {author} {\bibfnamefont
  {C.}~\bibnamefont {Berthier}}, \bibinfo {author} {\bibfnamefont {H.-R.}\
  \bibnamefont {Ott}}, \bibinfo {author} {\bibfnamefont {H.~M.}\ \bibnamefont
  {R\o{}nnow}}, \bibinfo {author} {\bibfnamefont {C.}~\bibnamefont {R\"uegg}},
  \ and\ \bibinfo {author} {\bibfnamefont {J.}~\bibnamefont {Mesot}},\ }\href
  {\doibase 10.1103/PhysRevLett.110.187201} {\bibfield  {journal} {\bibinfo
  {journal} {Phys. Rev. Lett.}\ }\textbf {\bibinfo {volume} {110}},\ \bibinfo
  {pages} {187201} (\bibinfo {year} {2013})}\BibitemShut {NoStop}%
\bibitem [{\citenamefont {Mourigal}\ \emph {et~al.}(2013)\citenamefont
  {Mourigal}, \citenamefont {Enderle}, \citenamefont {Kl{\"o}pperpieper},
  \citenamefont {Caux}, \citenamefont {Stunault},\ and\ \citenamefont
  {R{\o}nnow}}]{Mourigal2013}%
  \BibitemOpen
  \bibfield  {author} {\bibinfo {author} {\bibfnamefont {M.}~\bibnamefont
  {Mourigal}}, \bibinfo {author} {\bibfnamefont {M.}~\bibnamefont {Enderle}},
  \bibinfo {author} {\bibfnamefont {A.}~\bibnamefont {Kl{\"o}pperpieper}},
  \bibinfo {author} {\bibfnamefont {J.-S.}\ \bibnamefont {Caux}}, \bibinfo
  {author} {\bibfnamefont {A.}~\bibnamefont {Stunault}}, \ and\ \bibinfo
  {author} {\bibfnamefont {H.~M.}\ \bibnamefont {R{\o}nnow}},\ }\href {\doibase
  10.1038/nphys2652} {\bibfield  {journal} {\bibinfo  {journal} {Nature
  Physics}\ }\textbf {\bibinfo {volume} {9}},\ \bibinfo {pages} {435} (\bibinfo
  {year} {2013})}\BibitemShut {NoStop}%
\bibitem [{\citenamefont {Blosser}\ \emph {et~al.}(2017)\citenamefont
  {Blosser}, \citenamefont {Kestin}, \citenamefont {Povarov}, \citenamefont
  {Bewley}, \citenamefont {Coira}, \citenamefont {Giamarchi},\ and\
  \citenamefont {Zheludev}}]{Blosser2017}%
  \BibitemOpen
  \bibfield  {author} {\bibinfo {author} {\bibfnamefont {D.}~\bibnamefont
  {Blosser}}, \bibinfo {author} {\bibfnamefont {N.}~\bibnamefont {Kestin}},
  \bibinfo {author} {\bibfnamefont {K.~Y.}\ \bibnamefont {Povarov}}, \bibinfo
  {author} {\bibfnamefont {R.}~\bibnamefont {Bewley}}, \bibinfo {author}
  {\bibfnamefont {E.}~\bibnamefont {Coira}}, \bibinfo {author} {\bibfnamefont
  {T.}~\bibnamefont {Giamarchi}}, \ and\ \bibinfo {author} {\bibfnamefont
  {A.}~\bibnamefont {Zheludev}},\ }\href {\doibase 10.1103/PhysRevB.96.134406}
  {\bibfield  {journal} {\bibinfo  {journal} {Phys. Rev. B}\ }\textbf {\bibinfo
  {volume} {96}},\ \bibinfo {pages} {134406} (\bibinfo {year}
  {2017})}\BibitemShut {NoStop}%
\bibitem [{\citenamefont {Ward}\ \emph {et~al.}(2017)\citenamefont {Ward},
  \citenamefont {Mena}, \citenamefont {Bouillot}, \citenamefont {Kollath},
  \citenamefont {Giamarchi}, \citenamefont {Schmidt}, \citenamefont {Normand},
  \citenamefont {Kr\"amer}, \citenamefont {Biner}, \citenamefont {Bewley},
  \citenamefont {Guidi}, \citenamefont {Boehm}, \citenamefont {McMorrow},\ and\
  \citenamefont {R\"uegg}}]{Ward2017}%
  \BibitemOpen
  \bibfield  {author} {\bibinfo {author} {\bibfnamefont {S.}~\bibnamefont
  {Ward}}, \bibinfo {author} {\bibfnamefont {M.}~\bibnamefont {Mena}}, \bibinfo
  {author} {\bibfnamefont {P.}~\bibnamefont {Bouillot}}, \bibinfo {author}
  {\bibfnamefont {C.}~\bibnamefont {Kollath}}, \bibinfo {author} {\bibfnamefont
  {T.}~\bibnamefont {Giamarchi}}, \bibinfo {author} {\bibfnamefont {K.~P.}\
  \bibnamefont {Schmidt}}, \bibinfo {author} {\bibfnamefont {B.}~\bibnamefont
  {Normand}}, \bibinfo {author} {\bibfnamefont {K.~W.}\ \bibnamefont
  {Kr\"amer}}, \bibinfo {author} {\bibfnamefont {D.}~\bibnamefont {Biner}},
  \bibinfo {author} {\bibfnamefont {R.}~\bibnamefont {Bewley}}, \bibinfo
  {author} {\bibfnamefont {T.}~\bibnamefont {Guidi}}, \bibinfo {author}
  {\bibfnamefont {M.}~\bibnamefont {Boehm}}, \bibinfo {author} {\bibfnamefont
  {D.~F.}\ \bibnamefont {McMorrow}}, \ and\ \bibinfo {author} {\bibfnamefont
  {C.}~\bibnamefont {R\"uegg}},\ }\href {\doibase
  10.1103/PhysRevLett.118.177202} {\bibfield  {journal} {\bibinfo  {journal}
  {Phys. Rev. Lett.}\ }\textbf {\bibinfo {volume} {118}},\ \bibinfo {pages}
  {177202} (\bibinfo {year} {2017})}\BibitemShut {NoStop}%
\bibitem [{\citenamefont {Gannon}\ \emph {et~al.}(2019)\citenamefont {Gannon},
  \citenamefont {Zaliznyak}, \citenamefont {Wu}, \citenamefont {Feiguin},
  \citenamefont {Tsvelik}, \citenamefont {Demmel}, \citenamefont {Qiu},
  \citenamefont {Copley}, \citenamefont {Kim},\ and\ \citenamefont
  {Aronson}}]{Gannon2019}%
  \BibitemOpen
  \bibfield  {author} {\bibinfo {author} {\bibfnamefont {W.~J.}\ \bibnamefont
  {Gannon}}, \bibinfo {author} {\bibfnamefont {I.~A.}\ \bibnamefont
  {Zaliznyak}}, \bibinfo {author} {\bibfnamefont {L.~S.}\ \bibnamefont {Wu}},
  \bibinfo {author} {\bibfnamefont {A.~E.}\ \bibnamefont {Feiguin}}, \bibinfo
  {author} {\bibfnamefont {A.~M.}\ \bibnamefont {Tsvelik}}, \bibinfo {author}
  {\bibfnamefont {F.}~\bibnamefont {Demmel}}, \bibinfo {author} {\bibfnamefont
  {Y.}~\bibnamefont {Qiu}}, \bibinfo {author} {\bibfnamefont {J.~R.~D.}\
  \bibnamefont {Copley}}, \bibinfo {author} {\bibfnamefont {M.~S.}\
  \bibnamefont {Kim}}, \ and\ \bibinfo {author} {\bibfnamefont {M.~C.}\
  \bibnamefont {Aronson}},\ }\href {\doibase 10.1038/s41467-019-08715-y}
  {\bibfield  {journal} {\bibinfo  {journal} {Nature Communications}\ }\textbf
  {\bibinfo {volume} {10}},\ \bibinfo {pages} {1123} (\bibinfo {year}
  {2019})}\BibitemShut {NoStop}%
\bibitem [{\citenamefont {Yang}\ \emph {et~al.}(2019)\citenamefont {Yang},
  \citenamefont {Wu}, \citenamefont {Xu}, \citenamefont {Wang},\ and\
  \citenamefont {Wu}}]{Yang2019}%
  \BibitemOpen
  \bibfield  {author} {\bibinfo {author} {\bibfnamefont {W.}~\bibnamefont
  {Yang}}, \bibinfo {author} {\bibfnamefont {J.}~\bibnamefont {Wu}}, \bibinfo
  {author} {\bibfnamefont {S.}~\bibnamefont {Xu}}, \bibinfo {author}
  {\bibfnamefont {Z.}~\bibnamefont {Wang}}, \ and\ \bibinfo {author}
  {\bibfnamefont {C.}~\bibnamefont {Wu}},\ }\href {\doibase
  10.1103/PhysRevB.100.184406} {\bibfield  {journal} {\bibinfo  {journal}
  {Phys. Rev. B}\ }\textbf {\bibinfo {volume} {100}},\ \bibinfo {pages}
  {184406} (\bibinfo {year} {2019})}\BibitemShut {NoStop}%
\bibitem [{\citenamefont {Keselman}\ \emph {et~al.}(2020)\citenamefont
  {Keselman}, \citenamefont {Balents},\ and\ \citenamefont
  {Starykh}}]{keselman2020}%
  \BibitemOpen
  \bibfield  {author} {\bibinfo {author} {\bibfnamefont {A.}~\bibnamefont
  {Keselman}}, \bibinfo {author} {\bibfnamefont {L.}~\bibnamefont {Balents}}, \
  and\ \bibinfo {author} {\bibfnamefont {O.~A.}\ \bibnamefont {Starykh}},\
  }\href@noop {} {\enquote {\bibinfo {title} {Dynamical signatures of
  quasiparticle interactions in quantum spin chains},}\ } (\bibinfo {year}
  {2020}),\ \Eprint {http://arxiv.org/abs/2005.12399} {arXiv:2005.12399
  [cond-mat.str-el]} \BibitemShut {NoStop}%
\bibitem [{\citenamefont {Essler}\ \emph {et~al.}(2010)\citenamefont {Essler},
  \citenamefont {Frahm}, \citenamefont {G\"ohmann}, \citenamefont {Kl\"umper},\
  and\ \citenamefont {Korepin}}]{EsslerBook}%
  \BibitemOpen
  \bibfield  {author} {\bibinfo {author} {\bibfnamefont {F.}~\bibnamefont
  {Essler}}, \bibinfo {author} {\bibfnamefont {H.}~\bibnamefont {Frahm}},
  \bibinfo {author} {\bibfnamefont {F.}~\bibnamefont {G\"ohmann}}, \bibinfo
  {author} {\bibfnamefont {A.}~\bibnamefont {Kl\"umper}}, \ and\ \bibinfo
  {author} {\bibfnamefont {V.~E.}\ \bibnamefont {Korepin}},\ }\href@noop {}
  {\emph {\bibinfo {title} {The One-Dimensional Hubbard Model}}}\ (\bibinfo
  {publisher} {Cambridge University Press},\ \bibinfo {address} {Cambridge,
  England},\ \bibinfo {year} {2010})\BibitemShut {NoStop}%
\bibitem [{\citenamefont {Kohno}(2010)}]{Kohno2010}%
  \BibitemOpen
  \bibfield  {author} {\bibinfo {author} {\bibfnamefont {M.}~\bibnamefont
  {Kohno}},\ }\href@noop {} {\bibfield  {journal} {\bibinfo  {journal} {Phys.
  Rev. Lett.}\ }\textbf {\bibinfo {volume} {105}},\ \bibinfo {pages} {106402}
  (\bibinfo {year} {2010})}\BibitemShut {NoStop}%
\bibitem [{\citenamefont {Kohno}(2012)}]{Kohno2012}%
  \BibitemOpen
  \bibfield  {author} {\bibinfo {author} {\bibfnamefont {M.}~\bibnamefont
  {Kohno}},\ }\href@noop {} {\bibfield  {journal} {\bibinfo  {journal} {Phys.
  Rev. Lett.}\ }\textbf {\bibinfo {volume} {108}},\ \bibinfo {pages} {076401}
  (\bibinfo {year} {2012})}\BibitemShut {NoStop}%
\bibitem [{\citenamefont {Nocera}\ \emph {et~al.}(2018)\citenamefont {Nocera},
  \citenamefont {Essler},\ and\ \citenamefont {Feiguin}}]{Nocera2018}%
  \BibitemOpen
  \bibfield  {author} {\bibinfo {author} {\bibfnamefont {A.}~\bibnamefont
  {Nocera}}, \bibinfo {author} {\bibfnamefont {F.~H.~L.}\ \bibnamefont
  {Essler}}, \ and\ \bibinfo {author} {\bibfnamefont {A.~E.}\ \bibnamefont
  {Feiguin}},\ }\href {\doibase 10.1103/PhysRevB.97.045146} {\bibfield
  {journal} {\bibinfo  {journal} {Phys. Rev. B}\ }\textbf {\bibinfo {volume}
  {97}},\ \bibinfo {pages} {045146} (\bibinfo {year} {2018})}\BibitemShut
  {NoStop}%
\bibitem [{\citenamefont {Yang}\ and\ \citenamefont
  {Feiguin}(2016)}]{Yang2016}%
  \BibitemOpen
  \bibfield  {author} {\bibinfo {author} {\bibfnamefont {C.}~\bibnamefont
  {Yang}}\ and\ \bibinfo {author} {\bibfnamefont {A.~E.}\ \bibnamefont
  {Feiguin}},\ }\href@noop {} {\bibfield  {journal} {\bibinfo  {journal} {Phys.
  Rev. B}\ }\textbf {\bibinfo {volume} {93}},\ \bibinfo {pages} {081107}
  (\bibinfo {year} {2016})}\BibitemShut {NoStop}%
\bibitem [{\citenamefont {Kidd}\ \emph {et~al.}(2008)\citenamefont {Kidd},
  \citenamefont {Valla}, \citenamefont {Johnson}, \citenamefont {Kim},
  \citenamefont {Gu},\ and\ \citenamefont {Homes}}]{kidd08}%
  \BibitemOpen
  \bibfield  {author} {\bibinfo {author} {\bibfnamefont {T.~E.}\ \bibnamefont
  {Kidd}}, \bibinfo {author} {\bibfnamefont {T.}~\bibnamefont {Valla}},
  \bibinfo {author} {\bibfnamefont {P.~D.}\ \bibnamefont {Johnson}}, \bibinfo
  {author} {\bibfnamefont {K.~W.}\ \bibnamefont {Kim}}, \bibinfo {author}
  {\bibfnamefont {G.~D.}\ \bibnamefont {Gu}}, \ and\ \bibinfo {author}
  {\bibfnamefont {C.~C.}\ \bibnamefont {Homes}},\ }\href {\doibase
  10.1103/PhysRevB.77.054503} {\bibfield  {journal} {\bibinfo  {journal} {Phys.
  Rev. B}\ }\textbf {\bibinfo {volume} {77}},\ \bibinfo {pages} {054503}
  (\bibinfo {year} {2008})}\BibitemShut {NoStop}%
\bibitem [{\citenamefont {Matsueda}\ \emph {et~al.}(2005)\citenamefont
  {Matsueda}, \citenamefont {Bulut}, \citenamefont {Tohyama},\ and\
  \citenamefont {Maekawa}}]{Matsueda2005}%
  \BibitemOpen
  \bibfield  {author} {\bibinfo {author} {\bibfnamefont {H.}~\bibnamefont
  {Matsueda}}, \bibinfo {author} {\bibfnamefont {N.}~\bibnamefont {Bulut}},
  \bibinfo {author} {\bibfnamefont {T.}~\bibnamefont {Tohyama}}, \ and\
  \bibinfo {author} {\bibfnamefont {S.}~\bibnamefont {Maekawa}},\ }\href@noop
  {} {\bibfield  {journal} {\bibinfo  {journal} {Phys. Rev. B}\ }\textbf
  {\bibinfo {volume} {72}},\ \bibinfo {pages} {075136} (\bibinfo {year}
  {2005})}\BibitemShut {NoStop}%
\bibitem [{\citenamefont {Zemljic}\ \emph {et~al.}(2008)\citenamefont
  {Zemljic}, \citenamefont {Prelovsek},\ and\ \citenamefont
  {Tohyama}}]{Zemljic2008}%
  \BibitemOpen
  \bibfield  {author} {\bibinfo {author} {\bibfnamefont {M.~M.}\ \bibnamefont
  {Zemljic}}, \bibinfo {author} {\bibfnamefont {P.}~\bibnamefont {Prelovsek}},
  \ and\ \bibinfo {author} {\bibfnamefont {T.}~\bibnamefont {Tohyama}},\
  }\href@noop {} {\bibfield  {journal} {\bibinfo  {journal} {Phys. Rev. Lett.}\
  }\textbf {\bibinfo {volume} {100}},\ \bibinfo {pages} {036402} (\bibinfo
  {year} {2008})}\BibitemShut {NoStop}%
\bibitem [{\citenamefont {Eckstein}\ and\ \citenamefont
  {Werner}(2013)}]{Eckstein2013}%
  \BibitemOpen
  \bibfield  {author} {\bibinfo {author} {\bibfnamefont {M.}~\bibnamefont
  {Eckstein}}\ and\ \bibinfo {author} {\bibfnamefont {P.}~\bibnamefont
  {Werner}},\ }\href {\doibase 10.1103/PhysRevLett.110.126401} {\bibfield
  {journal} {\bibinfo  {journal} {Phys. Rev. Lett.}\ }\textbf {\bibinfo
  {volume} {110}},\ \bibinfo {pages} {126401} (\bibinfo {year}
  {2013})}\BibitemShut {NoStop}%
\bibitem [{\citenamefont {Eckstein}\ and\ \citenamefont
  {Werner}(2014)}]{Eckstein2014}%
  \BibitemOpen
  \bibfield  {author} {\bibinfo {author} {\bibfnamefont {M.}~\bibnamefont
  {Eckstein}}\ and\ \bibinfo {author} {\bibfnamefont {P.}~\bibnamefont
  {Werner}},\ }\href {\doibase 10.1103/PhysRevLett.113.076405} {\bibfield
  {journal} {\bibinfo  {journal} {Phys. Rev. Lett.}\ }\textbf {\bibinfo
  {volume} {113}},\ \bibinfo {pages} {076405} (\bibinfo {year}
  {2014})}\BibitemShut {NoStop}%
\bibitem [{\citenamefont {Balzer}\ \emph {et~al.}(2015)\citenamefont {Balzer},
  \citenamefont {Wolf}, \citenamefont {McCulloch}, \citenamefont {Werner},\
  and\ \citenamefont {Eckstein}}]{Belzer2015}%
  \BibitemOpen
  \bibfield  {author} {\bibinfo {author} {\bibfnamefont {K.}~\bibnamefont
  {Balzer}}, \bibinfo {author} {\bibfnamefont {F.~A.}\ \bibnamefont {Wolf}},
  \bibinfo {author} {\bibfnamefont {I.~P.}\ \bibnamefont {McCulloch}}, \bibinfo
  {author} {\bibfnamefont {P.}~\bibnamefont {Werner}}, \ and\ \bibinfo {author}
  {\bibfnamefont {M.}~\bibnamefont {Eckstein}},\ }\href {\doibase
  10.1103/PhysRevX.5.031039} {\bibfield  {journal} {\bibinfo  {journal} {Phys.
  Rev. X}\ }\textbf {\bibinfo {volume} {5}},\ \bibinfo {pages} {031039}
  (\bibinfo {year} {2015})}\BibitemShut {NoStop}%
\bibitem [{\citenamefont {Gole\ifmmode~\check{z}\else \v{z}\fi{}}\ \emph
  {et~al.}(2015)\citenamefont {Gole\ifmmode~\check{z}\else \v{z}\fi{}},
  \citenamefont {Eckstein},\ and\ \citenamefont {Werner}}]{Golez2015}%
  \BibitemOpen
  \bibfield  {author} {\bibinfo {author} {\bibfnamefont {D.}~\bibnamefont
  {Gole\ifmmode~\check{z}\else \v{z}\fi{}}}, \bibinfo {author} {\bibfnamefont
  {M.}~\bibnamefont {Eckstein}}, \ and\ \bibinfo {author} {\bibfnamefont
  {P.}~\bibnamefont {Werner}},\ }\href {\doibase 10.1103/PhysRevB.92.195123}
  {\bibfield  {journal} {\bibinfo  {journal} {Phys. Rev. B}\ }\textbf {\bibinfo
  {volume} {92}},\ \bibinfo {pages} {195123} (\bibinfo {year}
  {2015})}\BibitemShut {NoStop}%
\bibitem [{\citenamefont {Bittner}\ \emph {et~al.}(2018)\citenamefont
  {Bittner}, \citenamefont {Gole\ifmmode~\check{z}\else \v{z}\fi{}},
  \citenamefont {Strand}, \citenamefont {Eckstein},\ and\ \citenamefont
  {Werner}}]{Bittner2018}%
  \BibitemOpen
  \bibfield  {author} {\bibinfo {author} {\bibfnamefont {N.}~\bibnamefont
  {Bittner}}, \bibinfo {author} {\bibfnamefont {D.}~\bibnamefont
  {Gole\ifmmode~\check{z}\else \v{z}\fi{}}}, \bibinfo {author} {\bibfnamefont
  {H.~U.~R.}\ \bibnamefont {Strand}}, \bibinfo {author} {\bibfnamefont
  {M.}~\bibnamefont {Eckstein}}, \ and\ \bibinfo {author} {\bibfnamefont
  {P.}~\bibnamefont {Werner}},\ }\href {\doibase 10.1103/PhysRevB.97.235125}
  {\bibfield  {journal} {\bibinfo  {journal} {Phys. Rev. B}\ }\textbf {\bibinfo
  {volume} {97}},\ \bibinfo {pages} {235125} (\bibinfo {year}
  {2018})}\BibitemShut {NoStop}%
\bibitem [{\citenamefont {Raczkowski}\ \emph {et~al.}(2015)\citenamefont
  {Raczkowski}, \citenamefont {Assaad},\ and\ \citenamefont
  {Pollet}}]{Raczkowski2015}%
  \BibitemOpen
  \bibfield  {author} {\bibinfo {author} {\bibfnamefont {M.}~\bibnamefont
  {Raczkowski}}, \bibinfo {author} {\bibfnamefont {F.~F.}\ \bibnamefont
  {Assaad}}, \ and\ \bibinfo {author} {\bibfnamefont {L.}~\bibnamefont
  {Pollet}},\ }\href {\doibase 10.1103/PhysRevB.91.045137} {\bibfield
  {journal} {\bibinfo  {journal} {Phys. Rev. B}\ }\textbf {\bibinfo {volume}
  {91}},\ \bibinfo {pages} {045137} (\bibinfo {year} {2015})}\BibitemShut
  {NoStop}%
\end{thebibliography}%
\end{document}